\documentclass{aastex62}

\usepackage{graphicx}
\usepackage{graphics}
\usepackage{amssymb}
\usepackage{bm}
\usepackage{times}

\newcommand{\aco}{\alpha_\mathrm{CO}}

\newcommand{\beq}{\begin{equation}}
\newcommand{\eeq}{\end{equation}}

\newcommand{\cm}{cm$^{-2}$}

\newcommand{\Msun}{{\rm M{_\odot}}}
\newcommand{\kms}{km~s$^{-1}$}

\newcommand{\Mmol}{{\rm M_{mol}}}

\newcommand{\hi}{H{\sc i}}

\newcommand{\ci}{[C{\sc ii}]}
\newcommand{\cii}{[C{\sc ii}]\,158$\mu$m}

\newcommand{\zcii}{z_{\scriptsize \textrm{[C} \textsc{ii} \textrm{]} }}
\newcommand{\lcii}{L_{\scriptsize \textrm{[C} \textsc{ii} \textrm{]} }}

\shorttitle{The nature of H{\sc i}-selected galaxies at $z \approx 4$}
\shortauthors{Kaur et al.}
\begin{document}
\title{The nature of H{\sc  i}-absorption-selected galaxies at $z \approx 4$}

\correspondingauthor{Nissim Kanekar}
\email{nkanekar@ncra.tifr.res.in}

\author{B. Kaur} 
\affiliation{National Centre for Radio Astrophysics, Tata Institute of Fundamental Research, 
Pune University, Pune 411007, India}

\author{N. Kanekar} 
\affiliation{National Centre for Radio Astrophysics, Tata Institute of Fundamental Research, 
Pune University, Pune 411007, India}

\author{M. Rafelski} 
\affiliation{Space Telescope Science Institute, 3700 San Martin Drive, Baltimore, MD 21218, USA}
\affiliation{Department of Physics \& Astronomy, Johns Hopkins University, Baltimore, MD 21218, USA}

\author{M. Neeleman}
\affiliation{Max-Planck-Institut für Astronomie, Königstuhl 17, D-69117, Heidelberg, Germany}

\author{M. Revalski}
\affiliation{Space Telescope Science Institute, 3700 San Martin Drive, Baltimore, MD 21218, USA}

\author{J. X. Prochaska}
\affiliation{Department of Astronomy \& Astrophysics, UCO/Lick Observatory, University of California, 1156 High Street, Santa Cruz, CA 95064, USA}
\affiliation{Kavli Institute for the Physics and Mathematics of the Universe (Kavli IPMU), 5-1-5 Kashiwanoha, Kashiwa, 277-8583, Japan}

\begin{abstract}
We report a Karl G. Jansky Very Large Array (JVLA) search for redshifted CO(1--0) or CO(2--1) emission, and a Hubble Space Telescope Wide Field Camera~3 (HST-WFC3) search for rest-frame near-ultraviolet (NUV) stellar emission, from seven H{\sc i}-selected galaxies associated with high-metallicity ([M/H]~$\geq -1.3$) damped Ly$\alpha$ absorbers (DLAs) at $z \approx 4$. The galaxies were earlier identified by Atacama Large Millimeter/submillimeter Array imaging of their [C{\sc ii}]~158$\mu$m emission. We also used the JVLA to search for CO(2--1) emission from the field of a low-metallicity ([M/H]~$=-2.47$) DLA at $z \approx 4.8$. No statistically significant CO emission is detected from any of the galaxies, yielding upper limits of $\rm M_{mol} < (7.4 - 17.9) \times 10^{10} \times (\alpha_{CO}/4.36) \; M_\odot$ on their molecular gas mass. We detect rest-frame NUV emission from four of the seven [C{\sc ii}]~158$\mu$m-emitting galaxies, the first detections of the stellar continuum from H{\sc i}-selected galaxies at $z \gtrsim 4$. The HST-WFC3 images yield typical sizes of the stellar continua of $\approx 2-4$~kpc and  inferred dust-unobscured star-formation rates (SFRs) of $\approx 5.0-17.5 \ \rm M_\odot$~yr$^{-1}$, consistent with, or slightly lower than, the total SFRs estimated from the far-infrared (FIR) luminosity.  We further stacked the CO(2--1) emission signals of six [C{\sc ii}]~158$\mu$m-emitting galaxies in the image plane. Our non-detection of CO(2--1) emission in the stacked image yields the limit $\rm M_{mol} < 4.1 \times 10^{10} \times (\alpha_{CO}/4.36) \; M_\odot$ on the average molecular gas mass of the six galaxies. Our molecular gas mass estimates and NUV SFR estimates in H{\sc i}-selected galaxies at $z \approx 4$ are consistent with those of main-sequence galaxies with similar [C{\sc ii}]~158$\mu$m and FIR luminosities at similar redshifts. However, the NUV emission in the H{\sc i}-selected galaxies appears more extended than that in main-sequence galaxies at similar redshifts. 
\end{abstract}


\section{Introduction}

Our current picture of galaxy evolution is largely based on galaxies detected in emission in deep optical or 
near-infrared images \citep[e.g.][]{Madau14}. While much progress has been made in understanding the nature of high-redshift galaxies and their evolution using galaxy samples selected based on their stellar properties, such samples contain an inherent bias towards the brighter end of the luminosity distribution. This luminosity bias can be eliminated by selecting galaxies via their absorption signatures in the spectra of background QSOs. The best such   signature for carrying out a census of $z>2$ galaxies is the damping wing of the Ly$\alpha$ absorption line at high \hi\ column densities. For more than 30~years \citep{Wolfe86}, the highest \hi\ column density systems ($\rm N_{HI} \geq 2 \times 10^{20}$~\cm) in QSO spectra, the damped Ly$\alpha$ absorbers (DLAs), have been used as a signpost for the presence of a galaxy along, or close to, the QSO sightline. These absorption-selected galaxies represent typical members of the galaxy population at a given redshift, yielding a less biased sampling of the galaxy luminosity distribution 
\citep[e.g.][]{Gardner97,Pontzen08,Cen12,Kanekar20}.

QSO absorption spectroscopy allows both the identification of DLAs and the characterization of the physical properties of gas along the QSO sightline, including its absorption kinematics, metallicity, molecular fraction, number density, pressure, temperature, etc \citep[e.g.][]{Prochaska97,Noterdaeme08,Neeleman13,Neeleman15,Kanekar14,Balashev17,Klimenko20}. Such absorption studies have shown that the average DLA metallicity increases with decreasing redshift \citep{Rafelski12,Poudel20}, consistent with increased metal enrichment due to star formation  \citep[although with lower enrichment than expected; e.g. ][]{Peroux20}. However, to understand the nature of the galaxies that give rise to DLAs, and connect them to emission-selected samples at similar redshifts, it is critical to detect the DLA galaxies (i.e. \hi-selected galaxies) in emission. Unfortunately, the faintness of the absorbing galaxy in the optical waveband, relative to the background QSO, has meant that only $\approx 20$ \hi-selected galaxies have so far been identified at $z \gtrsim 2$  \citep[e.g.,][]{Fumagalli15,Krogager17,Ranjan20,Rhodin20}, and with a complex selection function. 

Recently, significant progress has been made in identifying the galaxies associated with high-$z$ DLAs, using mm-wave studies with the Atacama Large Millimeter/submillimeter Array (ALMA). ALMA searches for redshifted \cii\ and CO emission associated with high-metallicity DLAs have resulted in the identification of more than ten H{\sc i}-selected galaxies, out to $z \approx 4.5$ \citep[e.g.,][]{Neeleman17,Neeleman18,Neeleman19,Fynbo18,Kanekar20}. The ALMA DLA galaxies have been found to lie mostly at large impact parameters to the QSO sightline, $b \gtrsim 15$~kpc, very different from the \hi-selected galaxies identified in optical imaging \citep[e.g.,][]{Neeleman19,Kanekar20}.
The large impact parameters suggest that the DLAs may arise in gas clumps in the circumgalactic medium (CGM) of the \hi-selected galaxies, rather than in the galaxy disks themselves. This would require high \hi-column density gas to be present in the CGM of \hi-selected galaxies, within tens of kpc, unlike the case at low redshifts \citep[e.g.][]{Werk2014}. Follow-up ALMA \cii\ mapping studies of two of the \hi-selected galaxies have revealed a cold and rotating disk galaxy (the ``Wolfe disk'') at $z \approx 4.26$ \citep{Neeleman20} and a merging system of galaxies at $z \approx 3.80$ \citep{Prochaska19}. 

\begin{table}
\centering
\caption{The target DLAs of our sample, arranged by increasing DLA redshift. The columns are (1)~the QSO name, (2)~the QSO redshift, (3)~the DLA redshift, (4)~the logarithm of the \hi\ column density, (5)~the absorber metallicity, [M/H] \citep{Rafelski12,Rafelski14}, (6)~the \cii\ emission redshift, $\zcii$, (7)~~the J2000  coordinates of the \cii\ emission, and (8)~the DLA galaxy identifier. Note that \cii\ emission has not so far been detected from the galaxy associated with the $z = 4.8166$ DLA towards J1054+1633.
\label{tab:dla}   }
\vspace{0.1cm}    
\begin{tabular}{lccccccccc}
\hline
QSO & $z_{\rm QSO}$ & $z_{\rm abs}$  & log[N$_{\scriptsize \textrm{H}\textsc{i}}/{\rm cm^2}$] & [M/H] & $\zcii$ & J2000 co-ordinates & DLA galaxy \\
\hline

J1201+2117     & 4.579 & 3.7975 & $21.35 \pm 0.15$ & $-0.75 \pm 0.15$ & 3.7978 & 12:01:10.26 , +21:17:56.2 & DLA1201g1 \\
J1054+1633     & 5.187 & 4.1346 & $21.00 \pm 0.10$ & $-0.70 \pm 0.11$ & 4.1344 & 10:54:45.67 , +16:33:34.7 & DLA1054g1 \\
               &       &        &                  &                  & 4.1341 & 10:54:45.93 , +16:33:36.1 & DLA1054g2 \\
J1443+2724     & 4.443 & 4.2241 & $21.00 \pm 0.10$ & $-0.95 \pm 0.20$ & 4.2227 & 14:43:31.29 , +27:24:38.3 & DLA1443g1 \\
               &       &        &                  &                  & 4.2276 & 14:43:31.41 , +27:24:38.8 & DLA1443g2 \\
J0817+1351     & 4.398 & 4.2584 & $21.30 \pm 0.15$ & $-1.15 \pm 0.15$ & 4.2603 & 08:17:40.86 , +13:51:38.2 & DLA0817g1 \\
J1101+0531     & 4.986 & 4.3446 & $21.30 \pm 0.10$ & $-1.07 \pm 0.12$ & 4.2433 & 11:01:34.34 , +05:31:37.8 & DLA1101g1 \\
J0834+2140     & 4.521 & 4.3900 & $21.00 \pm 0.20$ & $-1.30 \pm 0.20$ & 4.3896 & 08:34:29.71 , +21:40:23.3 & DLA0834g1 \\
J1054+1633     & 5.187 & 4.8166 & $20.95 \pm 0.15$ & $-2.47 \pm 0.15$ & $-$ & $-$ \\
\hline
\end{tabular}
\end{table}

To further understand the nature of these \hi-selected galaxies and relate them to the emission-selected population, we have begun a multi-wavelength programme to characterize their physical properties. As part of this programme, we have used the Karl G. Jansky Very Large Array (JVLA) to carry out a search for redshifted CO emission from the sample of H{\sc i}-selected galaxies at $z \approx 4$  with ALMA detections of \cii\ emission \citep{Neeleman17,Neeleman19}, as well as Hubble Space Telescope (HST) Wide Field Camera~3 (WFC3) imaging of their rest-frame near-ultraviolet (NUV) stellar emission. Our initial observations resulted in the first detection of CO emission and stellar emission from an H{\sc i}-selected galaxy at $z \approx 4.26$ \citep{Neeleman20}. In this paper, we report results from a JVLA search for CO(2--1) or CO(1--0) emission, and an HST-WFC3 search for rest-frame NUV emission, from a sample of seven \cii-emitting H{\sc i}-selected galaxies at $z \approx 3.8-4.4$.\footnote{We will assume a standard flat $\Lambda$ cold dark matter cosmology with $\Omega_{\Lambda}$ = 0.685, $\Omega_{m}$ = 0.315, and H$_0$ = 67.4~km s$^{-1}$ Mpc$^{-1}$ \citep{Planck20}.}

\section{Observations and Data Analysis}
\label{sec:obs}

 The targets of our programme are the \cii-emitting galaxies identified in the fields of six DLAs at $z \gtrsim 4$ \citep{Neeleman17,Neeleman19}; the DLAs and their basic absorption properties \citep{Rafelski12,Rafelski14} are listed in Table~\ref{tab:dla}. The JVLA and HST-WFC3 observations of one of these systems,  at $z \approx 4.26$ towards J0817+1351, were presented by \citet{Neeleman20}, and will hence not be discussed here except for updates to the HST-WFC3 analysis and results. Further, two of the DLAs, towards J1054+1633 and J1443+2724, each have two associated \cii\ emitters, with redshifts close to the absorption redshift. Thus, including J0817+1351, there are eight \hi-selected, \cii-emitting galaxies in our sample, in six DLA fields. The last three columns of Table~\ref{tab:dla} list, respectively, the \cii\ emission redshifts, the J2000 co-ordinates, and the adopted DLA galaxy identifiers. In addition, one of the sightlines, towards J1054+1633, has a second DLA at $z \approx 4.8166$, which has not so far been detected in \cii\ emission. However, this was automatically covered in our JVLA CO spectroscopy and HST-WFC3 imaging, and is hence included in Table~\ref{tab:dla}.

\subsection{JVLA observations and data analysis}
\label{sec:jvla}

The JVLA CO observations of our five target DLA fields \citep[i.e. excluding J0817+1351;][]{Neeleman20} were carried out between September 2018 and December 2019, in the D-array (proposal IDs: VLA/18A-403, PI: M. Neeleman;  VLA/19B-271; PI: N. Kanekar). For four DLAs at $z \approx 4.13 - 4.39$, we used the Q-band receivers to search for CO(2--1) emission, redshifted to the frequency range $\approx 42.78-44.90$~GHz. However, in the case of DLA1201g1 at $z \approx 3.7978$, the redshifted CO(2--1) line frequency is $\approx 48$~GHz, where the Q-band sensitivity is relatively low, due to a high system temperature. We hence chose to use the K-band receivers to cover the CO(1--0) line of this system, redshifted to $\approx 24.03$~GHz. The total on-source times in each field were between 1.7 and 6.5 hours.

The JVLA observations used the WIDAR correlator as the backend, with two 1~GHz IF bands, each sub-divided into eight 128-MHz sub-bands, and two circular polarizations. One of the 128~MHz sub-bands was used to cover the redshifted CO line frequency for each DLA field; this sub-band was further sub-divided into either 128 or 512 channels, yielding a velocity resolution of $\approx 2-7$~\kms\ and a total velocity coverage of $\approx 900$~\kms\ (Q-band) or $\approx 1600$~\kms\ (K-band). The rest of the 128-MHz sub-bands were used to search for continuum emission from the DLA galaxy and the background QSO. For the sightline towards J1054+1633, which has a second DLA at $z \approx 4.8166$, we used one of the 128-MHz sub-bands of the second 1-GHz IF band to simultaneously cover the redshifted CO(2--1) line from the galaxy associated with this DLA. 

\begin{table}
\centering
\caption{Details of the JVLA observations and results, in order of increasing \cii\ emission redshift. The columns are (1)~the DLA galaxy identifier, (2)~the \cii\ emission redshift, $\zcii$, (3)~the targeted CO transition, (4)~the redshifted CO line frequency, in GHz, (5)~the on-source time, in hours, (6)~the synthesized beam, in $'' \times ''$, 
(7)~the RMS noise, in $\mu$Jy/Bm, at a velocity resolution of $200$~\kms,  (8)~the $3\sigma$ upper limit on the integrated CO line flux density, assuming a Gaussian line profile, with line FWHM~$=200$~\kms, and (9)~the $3\sigma$ upper limit on the {\it observed} CO line luminosity, $\rm L{ '_{CO,obs}}$.
\label{tab:obs}   }
\vspace{0.2cm}    
\begin{tabular}{lcccccccc}
\hline
Galaxy & $\zcii$ & CO line & $\nu_{\rm obs}$  & Time  & Beam & $\rm RMS_{CO}$ & $\int S_\mathrm{{CO}} \; dV$  &  $\rm L'_{CO,obs}$ \\
& & & GHz & hrs  & $''\times''$  & $\mu$Jy/Bm & Jy~\kms  & $10^{10}$~K~\kms~pc$^2$\\
\hline

DLA1201g1   & 3.7978 & 1--0 & 24.03 & 6.1 & $3.2 \times 3.1$ &  25 & $<0.016$ & $< 0.96$\\
DLA1054g2   & 4.1341 & 2--1 & 44.90 & 2.6 & $1.8 \times 1.6$ & 130 & $<0.083$ & $< 1.43$\\
DLA1054g1   & 4.1344 & 2--1 & 44.90 & 2.6 & $1.8 \times 1.6$ & 145 & $<0.092$ & $< 1.60$\\
DLA1443g1   & 4.2227 & 2--1 & 44.14 & 6.0 & $1.6 \times 1.4$ & 102 & $<0.065$ & $< 1.16$\\
DLA1443g2   & 4.2276 & 2--1 & 44.10 & 6.0 & $1.6 \times 1.4$ & 105 & $<0.067$ & $< 1.19$\\
DLA1101g1   & 4.3433 & 2--1 & 43.15 & 1.7 & $1.5 \times 1.4$ & 190 & $<0.12$ & $< 2.26$\\
DLA0834g1   & 4.3896 & 2--1 & 42.77 & 6.5 & $2.4 \times 2.0$ &  77 & $<0.049$ & $< 0.93$\\
J1054+1633$^a$ & 4.8166$^a$ & 2--1 & 39.63 & 2.6 & $2.0 \times 1.8$ & 90 & $<0.057$ & $< 1.25$\\    
\hline
\end{tabular}
\vskip 0.1in
$^a$\cii\ emission has so far not been detected from this field, at the DLA redshift. The table hence lists the QSO name and the DLA redshift.\\

\end{table}

 The data editing and calibration were carried out in the Astronomical Image Processing System package \citep[{\sc aips};][]{Greisen03}, following standard procedures. The visibilities were initially inspected to identify non-working and low-sensitivity antennas, which were edited out. For each observing run, we calibrated the antenna-based gains and bandpasses, and applied these to the data on the target source. In case of multiple observing runs on a single DLA field, the calibrated visibilities from the different epochs were combined into a single data set, using the task {\sc dbcon}. This data set was used to make a continuum image of each DLA field, using natural weighting to maximize the signal-to-noise ratio. No continuum emission was detected from any of the \hi-selected galaxies; further, in no case was the QSO continuum emission found to be sufficient for self-calibration.

For each DLA field, the combined visibility data set of the 128-MHz sub-band centered on the CO(2--1) or CO(1--0) line frequency was then imaged in the Common Astronomy Software Applications package \citep[{\sc casa};][]{CASA} to produce the final spectral cube, again using natural weighting. The cubes were made at velocity resolutions of $200 - 400$~\kms, similar to the velocity width of the \cii\ emission \citep{Neeleman17,Neeleman19}. The cubes were visually searched for CO emission at the locations of the ALMA \cii\ emission, and we also extracted spectra at these locations. 
Finally, for the DLA field at $z \approx 4.8166$ towards J1054+1633, which does not have a detection of \cii\ emission, we carried out a visual search of the cube for CO emission. 

The observational details and results are summarized in Table~\ref{tab:obs}, where the RMS noise values, and the $3\sigma$ upper limits to the CO line flux density and the line luminosity, are quoted at a velocity resolution of 200~\kms. We note that the JVLA synthesized beam has a full-width-at-half-maximum (FWHM)~$\geq 1\farcs4$ in all our target fields, corresponding to a transverse size $\gtrsim 10$~kpc at the DLA redshift. This is larger than both the spatial extent of the observed \cii-emission and the expected spatial extent of the CO emission \citep[e.g.][]{Daddi10,Kanekar20}. We therefore do not expect any emission to be resolved out.

\subsection{HST observations and data analysis}
\label{sec:hst}

\begin{figure}
\centering
\includegraphics[width=0.95\textwidth]{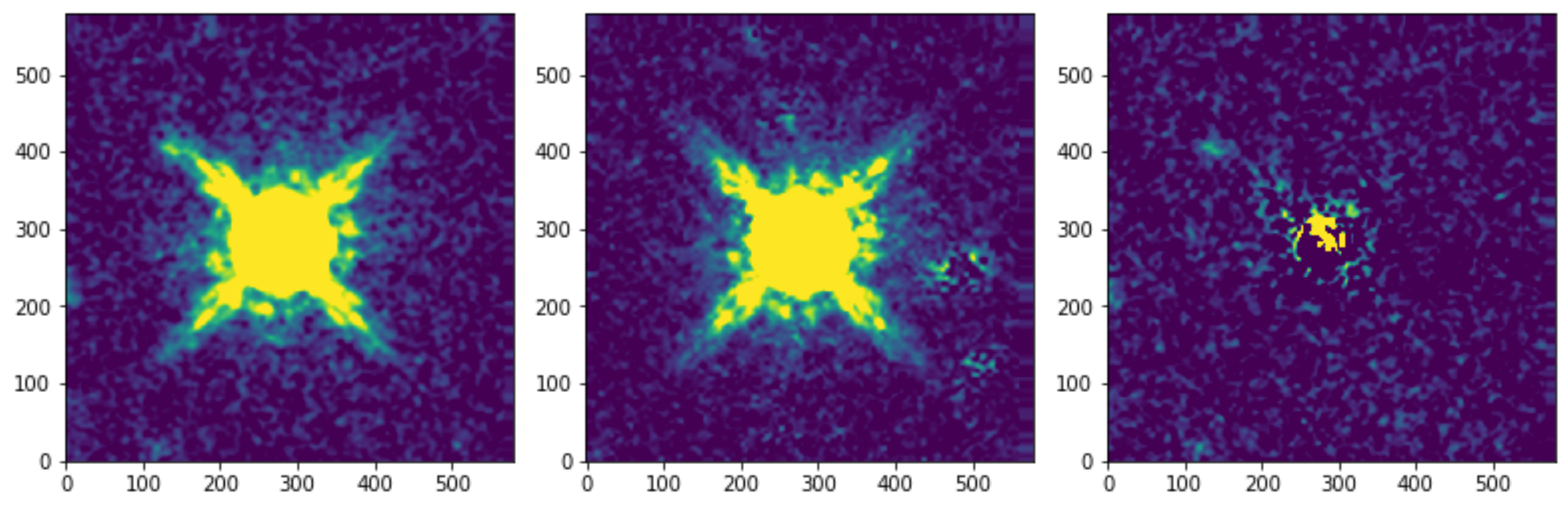}
\includegraphics[width=0.95\textwidth]{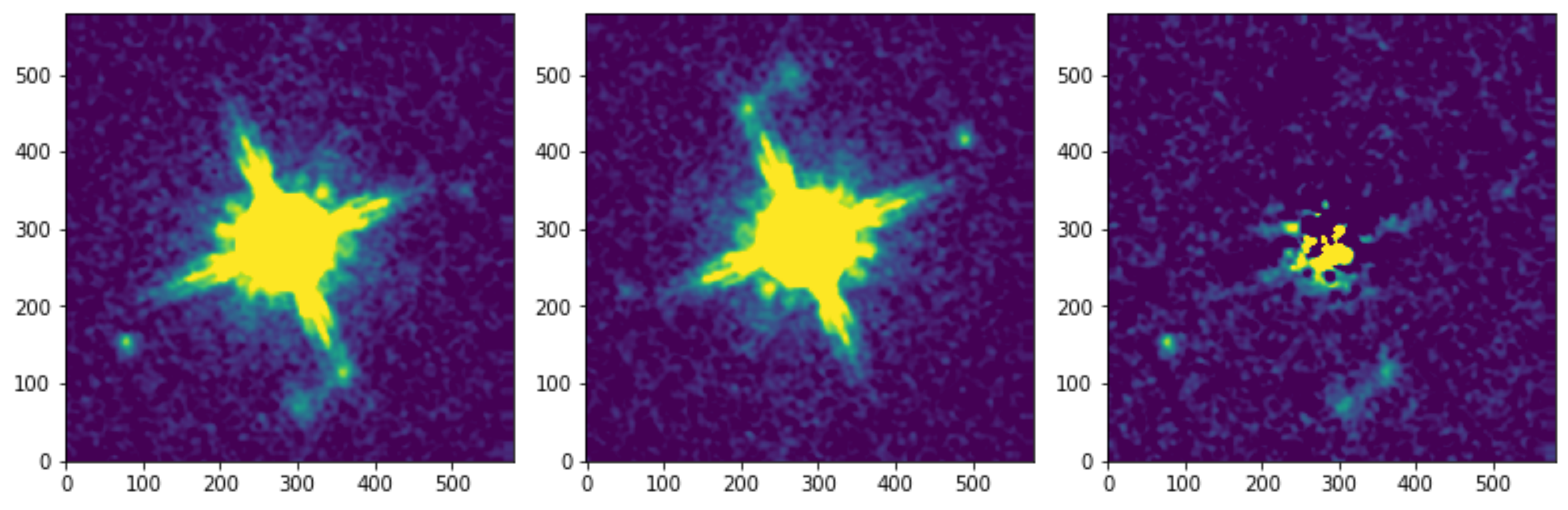}
\caption{Images showing the PSF-subtraction methodology for J1443+2724 (upper row) and J1201+2117 (bottom row), with the data (left), model (center) and residuals (right) shown for each field.  The contrast level is set to 1\% of the peak flux and the axes coordinates are in pixels after interpolation to 1/5th of the native pixel scale. The \cii-emitting galaxies of interest are located near the original locations of the upper-left and lower-right diffraction spikes for J1443+2724 and J1201+2117, respectively. The PSF model for J1443+2724 is based on the median of three bright stars, while the model for J1201+2117 is from the extracted quasar itself, rotated by 180$\degr$. }
\label{fig:psfsub}
\end{figure}

The HST-WFC3 observations of the five DLA fields were carried out in mid-2018 (programme ID, PID: 15410, PI: Neeleman), and late 2019 (PID: 15882, PI: Kanekar). We also re-analysed the HST-WFC3 data of the \hi-selected galaxy at $z \approx 4.2603$ towards J0817+1351, described in \citet{Neeleman20}. The F160W filter was used for all targets at $z > 4$, and the F140W filter for DLA1201g1, to cover the rest-frame NUV stellar continuum. Targets from PID 15410 were observed for one orbit using a standard IR-DITHER-BOX pattern, while targets from PID 15882 were observed for one orbit using a WIDE-7 pattern increased by a factor of 3 over that described in ISR 2016-14 to remove IR blobs and obtain cleaner images. All data were calibrated using the new IR filter-dependent sky flats \citep[WFC3 ISR 2021-01; ][]{Mack21}. 

For each field, image mosaics were created using AstroDrizzle, drizzled to a pixel scale of 0.06~mas, with TweakReg used to align the astrometry to the GAIA DR2 astrometry \citep{Gaia18}; this yields an absolute astrometric uncertainty of $\approx 0\farcs01$. The effective angular resolution of the HST images is $\approx 0\farcs3$, as determined from a Gaussian fit to the point spread function (PSF) of the quasars.

Rest-frame near-UV emission was detected from DLA1201g1, DLA1054g1, DLA1443g1, and DLA0834g1, 
in addition to the  emission from DLA0817g1 \citep{Neeleman20}. 
Emission was not detected from DLA1101g1, DLA1054g2, or DLA1443g2.

\begin{table}
\centering
\caption{Details of the HST-WFC3 observations and results, with the DLA fields arranged in order of increasing \cii\ emission redshift. The columns are (1)~the DLA galaxy name, (2)~the \cii\ emission redshift, (3)~the HST-WFC3 filter, (4)~the AB magnitude or limits on this quantity, (5)~for detections, the Kron radius, R$_{\rm Kron}$, in $''$, (6)~for detections, the half-light radius, R$_{1/2}$, in $''$, (7)~the Kron radius, r$_{\rm Kron}$, in kpc, (8)~the half-light radius, r$_{1/2}$, in kpc, and (9)~the SFR inferred from the rest-frame NUV continuum, or $2\sigma$ upper limits on this quantity.  
\label{tab:HST} } 
\begin{tabular} {|l|c|c|c|c|c|c|c|c|}
\hline
DLA galaxy                 & $\zcii$    &   Filter &     $\rm m_{AB}$  &  R$_{\rm Kron}$ &  R$_{1/2}$ & r$_{\rm Kron}$ &  r$_{1/2}$ & SFR$_{\rm NUV}$  \\
                     &        &                            &          &  $''$     &  $''$  & kpc & kpc    & $\Msun$~yr$^{-1}$ \\
\hline

DLA1201g1$^a$       & 3.7978 & F140W  & $24.35 \pm  0.09$ & 0.30   &   0.43       &  2.2  & 3.1 &$17.5 \pm  1.5$  \\
DLA1054g2           & 4.1341 & F160W  & $> 24.90 \ {\rm arcsec^{-2}}$ & $-$    &    $-$ &$-$&    $-$& $<12.0$ \\
DLA1054g1$^b$       & 4.1344 & F160W  & $24.94 \pm  0.15$ & 0.38   &    0.28      & 2.7  & 2.0 &$11.7 \pm  1.6$ \\ 
                    &        &        & $23.26 \pm  0.06$ & 0.39   &    0.45      & 2.7 & 3.1 &$54.1 \pm  3.0$ \\
DLA1443g1           & 4.2227 & F160W  & $25.89 \pm  0.13$ & 0.26   &    0.18   & 1.8 &  1.2  & $5.0 \pm  0.6$  \\
DLA1443g2           & 4.2276 & F160W  & $> 25.3 \ {\rm arcsec^{-2}}$         & $-$   & $-$  &$-$ & $-$        & $<8.6$     \\
DLA0817g1           & 4.2603 & F160W  & $25.34 \pm 0.19$   & 0.59    &  0.25  & 4.1 & 1.7 & $8.4 \pm 1.5$ \\
DLA1101g1           & 4.3433 & F160W  & $> 25.2 \ {\rm arcsec^{-2}}$ & $-$       &  $-$    & $-$       &  $-$ & $<10.2$   \\
DLA0834g1           & 4.3896 & F160W  & $25.33 \pm 0.15$  & 0.39      &  0.22    & 2.7 & 1.5 & $8.9 \pm 1.3$ \\

\hline

\end{tabular}
\vskip 0.05in
$^a$~For DLA1201g1, all estimates include contributions from both merging galaxies. \\
$^b$~DLA1054g1 has two NUV galaxies close to the \cii\ emission region; we list both of these in the table. We have tentatively identified the fainter NUV object as likely to be associated with the \cii\ emitter; see Section~\ref{sec:nuv} for discussion.
\end{table}

The source fluxes of  DLA1443g1 and DLA1201g1 were contaminated by instrumental diffraction spikes originating from the nearby, bright quasars. We corrected for this by modeling the PSF of each image and subtracting out the contaminating source. For the field of J1443+2724, we visually identified and extracted sub-images of three bright, relatively isolated stars. We normalized the peak flux of each star to unity, interpolated them to a 1/5th pixel grid, determined their centroids by fitting Moffat profiles, aligned the stars to a common centroid, and calculated a median image. We then aligned this PSF model with the contaminating quasar interpolated to the same sub-pixel grid, scaled the PSF flux to match that of the quasar, subtracted it out, and interpolated back to the native pixel scale to generate a PSF-subtracted image. In the case of the field of J1201+2117, there were insufficient stars to generate a median model, so we extracted the quasar itself, rotated it by 180$\degr$, and followed the same interpolation, alignment, and subtraction technique. The data, models, and residuals for these objects are shown in Figure~\ref{fig:psfsub}.

\begin{figure}
\centering
	\includegraphics[width=0.99\textwidth]{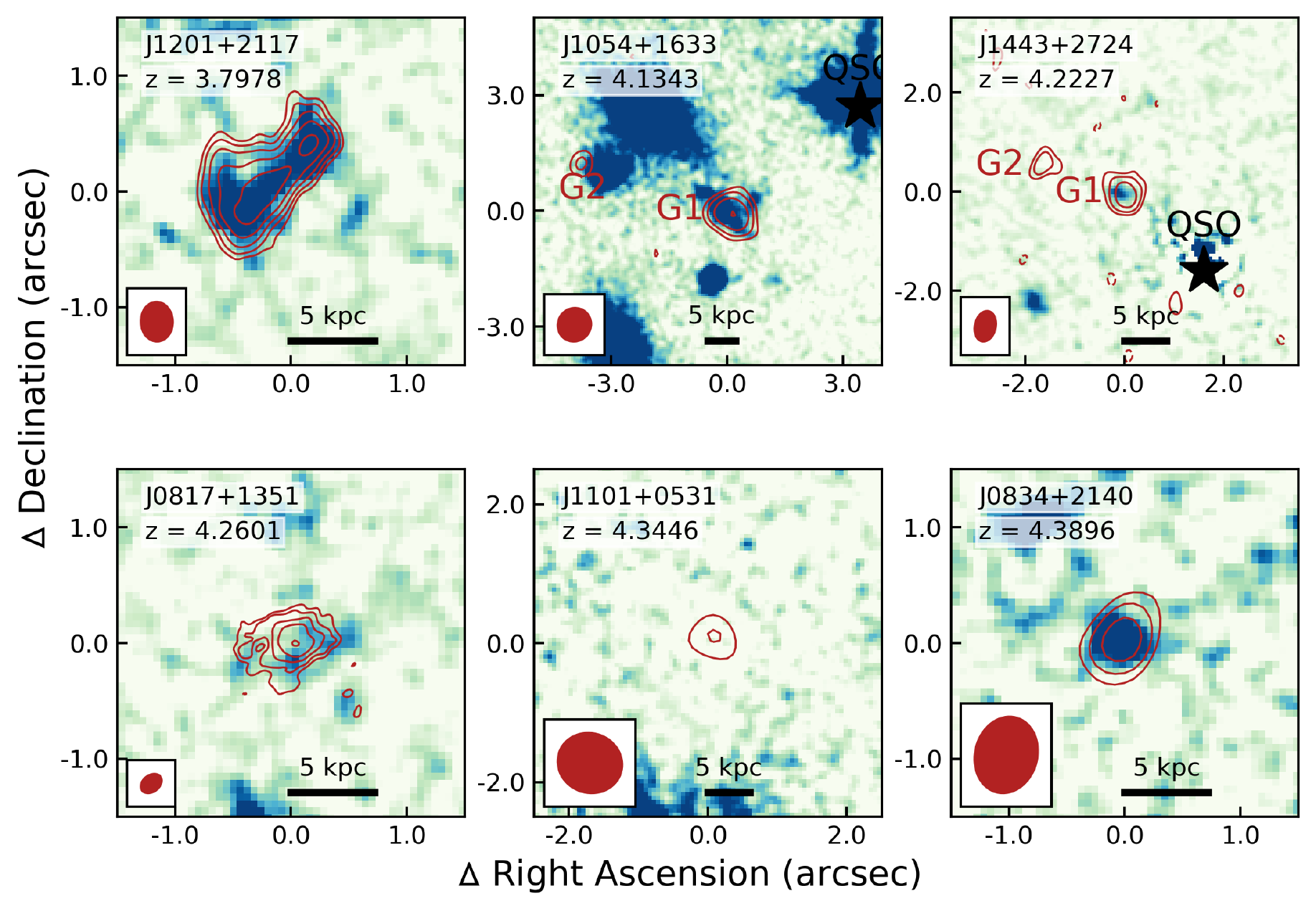}
\caption{HST-WFC3 NUV stellar continuum images of the six DLA fields (in colour scale), with the ALMA \cii\ emission overlaid (in contours). The images are centered on the \cii-emitting galaxies within the field, whose redshifts are indicated at the top left of each panel, below the QSO name. The fields of J1054+1633 and J1443+2724 each contain two \cii\ emitters; here, the images are centered on DLA1054g1 and DLA1443g1, respectively. For the fields of J1201+2117 and J1443+2714, we show the PSF-subtracted images. Contours mark the ALMA \cii\ emission, starting at $3\sigma$ and increasing in powers of $\sqrt{2}$. The resolution of the ALMA images is shown in the bottom-left inset of each panel.   
 \label{fig:HST}}
\end{figure}

To determine the photometry and the shape of the stellar continuum of the five galaxies (including DLA0817g1) with detected rest-frame NUV emission, we used Source Extractor~v.2.19.5 \citep{Bertin96}. We measured the total flux using Source Extractor’s flux$_{\rm auto}$, which gives the flux within an elliptical aperture with the Kron radius \citep{Kron80}. This yields total AB magnitudes of $\rm m_{AB}  \approx 24.3-25.9$ for the four new detections, with Kron radii of $0\farcs26-0\farcs59$. For the three non-detections, our $2\sigma$ limits are $\rm m_{AB} \gtrsim 25 \ {\rm arcsec^{-2}}$ . Half-light radii were also measured, yielding values of $0\farcs18-0\farcs45$; we note that the half-light radius for DLA1201g1 includes the light of both merging galaxies.  Fig.~\ref{fig:HST} shows the HST-WFC3 images of the DLA fields with the integrated \cii\ emission overlaid in contours \citep{Neeleman19,Prochaska19,Neeleman20}. Details of the HST-WFC3 observations and results are summarized in Table~\ref{tab:HST}.

We note that there are two NUV-emitting galaxies lying close to the \cii\ emission of DLA1054g1 in Fig.~\ref{fig:HST}. The current data do not allow us to formally identify which of these is associated with the \cii\ emitter, and we hence list details of both galaxies in Table~\ref{tab:HST}; see Section~\ref{sec:nuv} for discussion.

\begin{figure}
\centering
\includegraphics[width=0.99\textwidth]{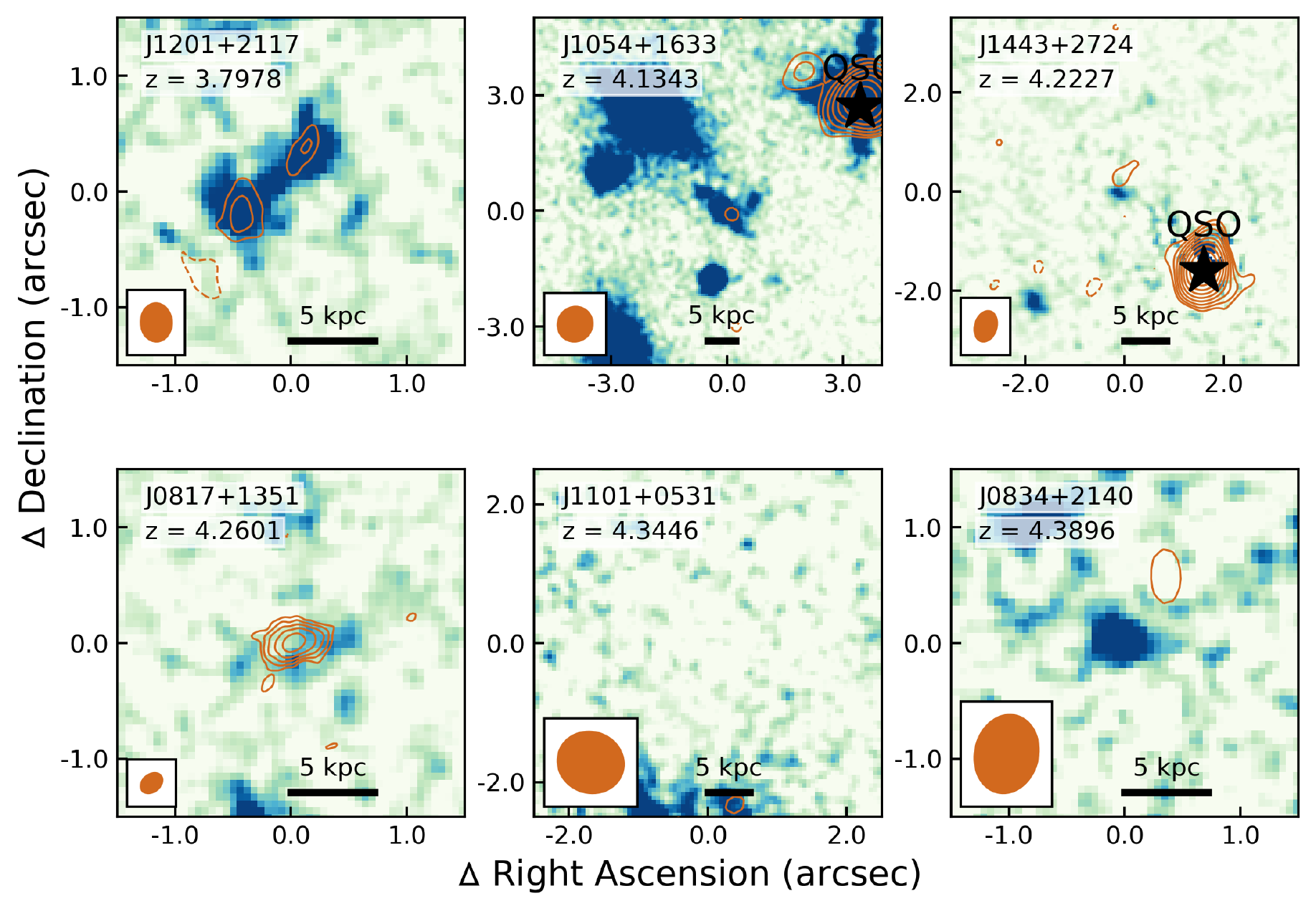}
\caption{HST-WFC3 NUV stellar continuum images of the six DLA fields, with ALMA dust emission overlaid (in contours). The images are centered on the \cii-emitting galaxies within the field, whose redshifts are indicated at the top left of each panel, below the QSO name. The fields of J1054+1633 and J1443+2724 contain two \cii\ emitters; here, the images are centered on DLA1054g1 and DLA1443g1, respectively. 
For J1201+2117 and J1443+2724, we show the PSF-subtracted images. Contours mark the ALMA dust emission, starting at $3\sigma$ and increasing in powers of $\sqrt{2}$. The resolution of the ALMA images is shown in the bottom-left inset.
\label{fig:HSTdust}}
\end{figure}

\section{Results}
\subsection{The CO line luminosity and the molecular gas mass}

Our JVLA observations find no evidence of CO(2--1) or CO(1--0) line emission from the $z \approx 4$ \hi-selected galaxies with \cii\ emission, or from the field of the $z \approx 4.8166$ DLA towards J1054+1633 (where no associated 
galaxy has been identified in emission), yielding upper limits on the velocity-integrated CO(2--1) or CO(1--0) line flux density. To convert these to limits on the {\it observed} CO line luminosity $\rm L'_{CO,obs}$ (in K~\kms~pc$^2$), we use the expression \citep{Solomon92}
\begin{equation}
        {\rm L'_{CO,obs}} = 3.25 \times 10^7 \times \frac{D_{L}^2}{(1+z)^3\nu_{\mathrm{CO}}^2}  \int S_\mathrm{CO} \; dV \; , 
\end{equation}
where $\int S_\mathrm{{CO}}~dV$ is the velocity-integrated line flux density in Jy~\kms, $D_{L}$ is the luminosity distance in Mpc, and $\nu_\mathrm{{CO}}$ is the redshifted CO line frequency, in GHz. Our $3\sigma$ upper limits on the  observed CO(2--1) or CO(1--0) line luminosity are listed in the last column of Table~\ref{tab:obs}. These must be corrected for the fact that the cosmic microwave background (CMB) temperature increases with increasing redshift, making it a brighter background against which the CO emission is being measured. Following \citet{dacunha13}, we use the following correction factor for the CMB to obtain the intrinsic CO line luminosity, $\rm L'_{CO}$
\begin{equation}
\frac{{\rm L'_{{CO,obs}}}}{{\rm L'_{{CO}}}} = 1 - \frac{B_{\nu}\left(T\mathrm{_{CMB}}(z)\right)}{B^J_{\nu}(T_{\mathrm{exc}})} \;, 
\end{equation}
where $T_\mathrm{{CMB}}$ is the CMB temperature at the DLA redshift, $T_{\mathrm{exc}}$ is the excitation temperature of the CO($J \rightarrow J-1$) line, and $B^J_{\nu}(T)$ is the black body intensity for the temperature $T$ and at the redshifted CO($J \rightarrow J-1$) line frequency. We assume a typical CO line excitation temperature of  $T_{\mathrm{exc}}\approx  35$~K, to find that the intrinsic CO line luminosity is higher than the observed luminosity by a factor of $\approx 1.5$ for all our targets, due to the above effect. 

In the case of DLA1201g1, at $z = 3.7978$, our limit on the CO(1--0) line luminosity $L'_\mathrm{CO(1-0)}$ can be directly used to infer an upper limit on the object's molecular gas mass, via the relation $\Mmol = \aco \times \rm L'_{CO(1-0)}$, where $\aco$ is the (assumed) CO-to-H$_2$ conversion factor \citep[e.g.][]{Carilli13,Bolatto13}. However, for the remaining galaxies, where we have searched for redshifted CO(2--1) emission, estimating the molecular gas mass also requires us to know the ratio of the intrinsic CO(2--1) and CO(1--0) line luminosities, $r_{21} \equiv \rm L'_{CO(2-1)}/L'_{CO(1-0)}$, which depends on physical conditions in the galaxy \citep[e.g.][]{Carilli13}. Earlier studies have obtained $r_{21}$ values of $\approx 0.76 - 0.92$ in both main-sequence galaxies  and sub-mm galaxies  \citep[e.g.][]{Aravena10,Bothwell12,Daddi15,Zavadsky15} at high redshifts. We will assume $r_{21} \approx 0.81$, obtained for normal star-forming galaxies at $z \approx 1.5-3$ \citep{Zavadsky15}.

Finally, the value of the CO-to-H$_2$ conversion factor, $\aco$, is known to depend inversely on galaxy metallicity, with low values, $\aco \lesssim 4.5 \; \Msun$~(K~\kms~pc$^2$)$^{-1}$, in high-metallicity galaxies and high values, $\aco \gtrsim 10 \; \Msun$~(K~\kms~pc$^2$)$^{-1}$, in low-metallicity galaxies \citep[e.g.][]{Bolatto13}. Most studies of high-redshift galaxies assume the Milky Way value, $\aco \approx 4.36 \; \Msun$~(K~\kms~pc$^2$)$^{-1}$ \citep[e.g.][]{Bolatto13,Tacconi20}. Unfortunately, we do not have measurements of the emission metallicities of our \hi-selected galaxies. Our absorption-based metallicities \citep{Neeleman17,Neeleman19} are measured at large impact parameters, $\approx 16-50$~kpc, from the galaxies, and thus only provide lower limits on the galaxy metallicity. However, \citet{Zanella18} has found evidence for a correlation between the ratio of the \cii\ line luminosity to the total SFR and the galaxy metallicity \citep[see Fig.~11 of ][]{Zanella18}. We use Equation~6 of \citet{Zanella18} in conjunction with our measured \cii\ line luminosities and SFR estimates  \citep[see Table~\ref{tab:derived}; ][]{Neeleman17,Neeleman19} to find that all eight of our \hi-selected galaxies have emission metallicities consistent with solar metallicity, within the $\approx 0.2$~dex spread of the \citet{Zanella18} relation. We further used these metallicities to infer the metallicity-dependent correction factor $\zeta(Z)$ \citep{Bolatto13,Tacconi20}, obtaining correction factors $\leq 25$\% for 7 of the 8 galaxies, and a weak upper limit to the correction factor in the last system \citep[where we only have an upper limit on its SFR, and hence a lower limit to the metallicity from the correlation of ][]{Zanella18}. Given the known uncertainties in the metallicity-dependent correction factor \citep{Bolatto13} and the fact that the metallicities of our galaxies are consistent within the errors with solar metallicity (the median correction factor is $\zeta(Z) \approx 1.0$), we will assume a single value of $\aco = 4.36 \; \Msun$~(K~\kms~pc$^2$)$^{-1}$ \citep{Tacconi20} for our sample of \hi-selected galaxies at $z \approx 4$. With these assumptions, our $3\sigma$ upper limits on the molecular gas mass of the \hi-selected galaxies lie in the range $\rm \Mmol < (7.4 - 17.9) \times 10^{10} \times (\aco/4.36) \; \Msun$.

\begin{figure}
\centering
\includegraphics [scale=0.34] {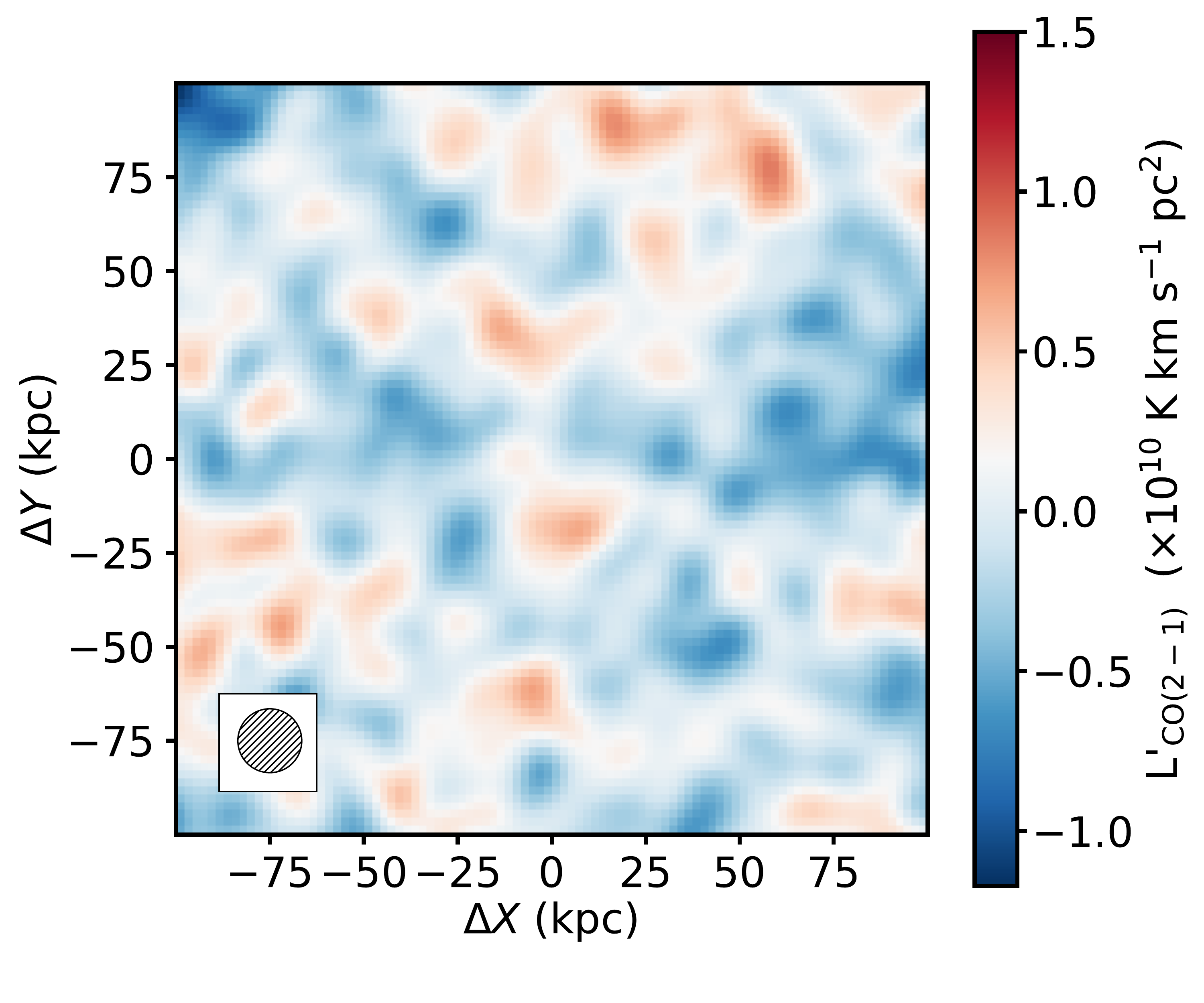}
\caption{Stacked CO(2--1) line luminosity image of six \hi-selected galaxies. The central pixel corresponds to the location of \cii\ line emission arising from the galaxies. The inset shows the resolution of the stacked image, 17~kpc. }
\label{fig:stack_image}
\end{figure}

\subsection{CO emission stacking: The average molecular gas mass}

For the detections of \cii\ line emission, we have accurate positions and redshifts for the \hi-selected galaxies \citep{Neeleman17,Neeleman19}. This information can be used to improve our sensitivity to the {\it average} CO line emission from our targets, by ``stacking'' the CO emission signals from the galaxies in the image plane, after aligning them in position and velocity \citep[see, e.g., ][for similar stacking analyses, applied to high-$z$ \cii\, CO, and H{\sc i} 21\,cm emission studies]{Zanella18,Coogan19,Chowdhury20}. We chose to restrict the stacking analysis to the CO(2--1) searches from the six \cii\ emitters of Table~\ref{tab:derived}, as the size of the synthesized beam for DLA1201g1, our only CO(1--0) search, is much larger than the sizes of the synthesized beams for the fields with CO(2--1) studies.

The stacking was carried out in the image plane, at a velocity resolution of $200$~\kms, centered on the \cii\ emission redshift of each galaxy. We first converted each 200~\kms-resolution spectral cube from flux density units to CO line luminosity units, and also corrected the CO line luminosity for CMB effects \citep{dacunha13}. 
Next, we smoothed each cube to the same spatial resolution (17~kpc), and re-sampled the smoothed cubes with spatial pixels of the same size (2~kpc). This was necessary because the cubes have different {\it spatial} resolutions at the different redshifted CO line frequencies as well as slightly different angular resolutions. For each \cii\ emitter, we then extracted a sub-cube of size 0.2~Mpc~$\times$~0.2~Mpc, centered at the location of the \cii\ line emission. Finally, we carried out a pixel-by-pixel weighted average of the 200~\kms\ image planes of the six \hi-selected galaxies centered on their redshifted CO line frequencies, using the inverse of the variance of the CO line luminosity in this plane as the weight for each galaxy (to maximize the signal-to-noise ratio), to obtain our final stacked CO(2--1) image.

Figure~\ref{fig:stack_image} shows our final stacked CO(2--1) line luminosity image; the axes are in units of kpc, and the image has a spatial resolution of 17~kpc.  We found no evidence of CO(2--1) emission in the stacked image, obtaining a $3\sigma$ upper limit of $\rm L'_{CO(2-1)} < 7.7 \times 10^{9}~$K~\kms~pc$^2$ on the average CO(2--1) line luminosity of the six \hi-selected galaxies. Again assuming $r_{21} = 0.81$ and $\aco = 4.36 \; \Msun$~(K~\kms~pc$^2$)$^{-1}$, this yields the 3$\sigma$ upper limit $\Mmol < 4.1 \times 10^{10} \, \times \, (\aco/4.36) \times (0.81/r_{21}) \; \Msun$ on the average molecular gas mass of the six \hi-selected galaxies at $z \approx 4$.

\subsection{Rest-frame NUV stellar continuum emission}
\label{sec:nuv}

Our HST-WFC3 observations have yielded new detections of the rest-frame NUV stellar continuum from four of our \hi-selected, \cii-emitting galaxies at $z \approx 4$, and upper limits on the NUV stellar emission for three systems. We have also used the new HST-WFC3 calibrations to improve the image of DLAJ0817g1 \citep{Neeleman20}. We use these NUV continuum measurements to infer the unobscured SFR of our galaxies\footnote{We will assume a Chabrier initial mass function \citep{Chabrier03} throughout this paper.}, using the local relation between NUV $2300$\AA\ emission and SFR \citep[e.g.][]{Kennicutt12}, assuming a flat spectrum in $L_\nu$ between $\approx 3000$\AA\ (the typical rest-frame wavelength of our data) and $2300$\AA\ \citep[e.g.][]{Kennicutt98}. This yields NUV SFRs of SFR$_{\rm NUV} \approx 5.0 - 17.5 \ \Msun$~yr$^{-1}$ for our detections, and $2\sigma$ upper limits of $< 8.6-12.0 \ \Msun$~yr$^{-1}$ for the non-detections, in all cases uncorrected for dust obscuration. Our NUV SFR estimates are listed in the last column of Table~\ref{tab:HST}. We note that our re-calibrated HST-WFC3 image at a smaller plate scale for J0817+1351 results in an NUV SFR about half as large as the previous estimate \citep{Neeleman20}. 

Of the two possible NUV counterparts of DLA1054g1, we tentatively identify the fainter of the two galaxies as likely to be associated with the \cii\ emitter. This is because its NUV SFR is $\approx 11.7 \ \Msun$~yr$^{-1}$, slightly lower than the total SFR of the galaxy estimated from the FIR continuum ($\approx 21 \ \Msun$~yr$^{-1}$), and thus consistent with dust obscuration of the NUV emission. The brighter of the two NUV galaxies would have a dust-unobscured SFR of $\approx 54 \ \Msun$~yr$^{-1}$ if located at $z = 4.1344$, i.e. higher than the total SFR estimate, and is thus unlikely to be the counterpart of DLA1054g1.

The stellar continuum sizes of the galaxies are quite compact, with half-light radii of $\approx 1.2-3.1$~kpc. This could be the result of the shallow depth of our HST observations, which may have missed faint extended emission. The sizes and shapes of the galaxies are very similar to the sizes and shapes of the \cii\ line emission (except possibly for DLAJ1443g1; see Fig.~\ref{fig:HST} and \citealt{Neeleman19}) suggesting that the NUV emission and the \cii\ emission are both tracing emission from the star-forming regions of the galaxies. This is particularly evident for both the cold disk galaxy DLA0817g1 \citep{Neeleman20} and  DLA1201g1, a pair of interacting galaxies discussed in \citet{Prochaska19}. For the latter system, the two galaxies and the \cii\ emission ``bridge'' between the galaxies can also be seen clearly in the HST-WFC3 image. 

In Figure \ref{fig:HSTdust}, we show the same HST images as in Figure \ref{fig:HST}, but with ALMA dust continuum contours overlaid. As with the \cii\ emission, the dust continuum broadly agrees in position with the rest-frame NUV stellar continuum, within the uncertainties of the observations. This is especially clear for the galaxies with high-significance detections of the dust continuum, i.e., DLA0817g1 and DLA1201g1.

\begin{table}
\centering
	\caption{The parameters of the \hi-selected galaxies, ordered by increasing \cii\ emission redshift \citep[][this work; Neeleman et al., in prep.]{Neeleman17,Neeleman19}. The columns are (1)~the DLA galaxy name, (2)~the \cii\ emission redshift, $\zcii$, (3)~the impact parameter to the QSO sightline, $b$, (4)~the \cii\ line luminosity, (5)~the FIR luminosity,  (6)~the $3\sigma$ JVLA upper limit on $\rm L'_{CO(1-0)}$, assuming $r_{21} = 0.81$, (7)~the $3\sigma$ JVLA upper limit on the molecular gas mass $\rm \Mmol$, assuming $\aco = 4.36 \; \Msun$~(K~\kms~pc$^2$)$^{-1}$, (8)~the SFR, from the FIR 160$\mu$m continuum luminosity \citep{Calzetti10}, for a Chabrier initial mass function, (9)~the SFR, from the rest-frame NUV measurements, (10)~the SFR, from the ALPINE \ci-SFR relation \citep{Schaerer20}, and (11)~the molecular gas depletion timescale, $\tau_{\rm depl}$. The last line of the table lists the $3\sigma$ limits on $\rm L'_{CO(1-0)}$ and $\rm \Mmol$ from the stacked CO(2--1) cube, along with the weighted-average of the redshift, the \cii\ line luminosity, the SFR$_{\rm FIR}$ and the SFR$_{\rm NUV}$, where the weights are the same as those used for the stacking of the CO(2-1) emission. For DLA0817g1, the \cii\ and CO line luminosities, and derived quantities, are from \citet{Neeleman17,Neeleman20}.
\label{tab:derived}}
\begin{tabular} {|l|c|c|c|c|c|c|c|c|c|c|}
\hline

	QSO & $\zcii$ & $b$ & $\rm \lcii$  & $\rm L_{FIR}$ &  $\rm L'_{CO(1-0)}$ & $\rm  \Mmol$  & SFR$_{\rm FIR}$ & SFR$_{\rm NUV}$ & SFR$_{\scriptsize \textrm{[C} \textsc{ii} \textrm{]}}$& $\tau_{\rm depl}$ \\ 

    &                  &    kpc  & $10^8 \ L_{\odot}$  &  $10^{10} \ L_{\odot}$& $10^{10}$ K~\kms~pc$^2$  & $10^{10} \; \Msun$ & $\Msun$~yr$^{-1}$ & $\Msun$~yr$^{-1}$ &  $\Msun$~yr$^{-1}$ &Gyr   \\ 

\hline
DLA1201g1           & 3.7978 & 18.1 & $8.7$  & $25$    & $< 1.8 $ & $< 7.8 $ & $25$   & $17.5$& 81 & $< 3.3$  \\  
DLA1054g2           & 4.1341 & 50.5 & $5.1 $ & $<15$   & $< 2.8 $ & $< 12.2$ & $< 17$ & $<12.0$& 48 & $ - $  \\   
DLA1054g1           & 4.1344 & 30.1 & $3.4 $ & $18$    & $< 2.6 $ & $< 11.3$ & $21$   & $11.7$&32 & $< 5.7$  \\   
DLA1443g1           & 4.2227 & 16.0 & $4.7$  & $14$    & $< 2.1 $ & $< 9.2$  & $16$   & $5.0$ &44 &$< 6.4$    \\    
DLA1443g2           & 4.2276 & 26.4 & $1.5$  & $< 7.2$ & $< 2.2 $ & $< 9.6$  & $< 7$   & $<8.6$& 14 & $-$  \\  
DLA0817g1$^\ast$    & 4.2603 & 42.9 & $30.2$ & $100$   & $ 2.9 $  & $12.8$   & $116$  & $8.4$&282  & $1.2$   \\      
DLA1101g1           & 4.3433 & 27.4 & $0.36$ & $< 7.0$ & $<4.1 $  & $< 17.9$ & $<7$   & $<10.2$& 3& $ - $   \\   
DLA0834g1           & 4.3896 & 27.3 & $1.0$  & $6.6$   & $< 1.7 $ & $< 7.4$  & $7$    & $8.9$& 9 & $< 10.6$  \\   
J1054+1633$^\dagger$ & 4.8166 & $-$ & $-$    & $-$     & $< 2.4 $ & $< 10.5$ & --     &  --  & -- & --    \\    
\hline                                                                                                          
CO(2--1) stack      & 4.2579  & $-$ & $2.6$  & $10$    & $< 0.95 $& $< 4.1$  & $12.1^\star$ &  $9.0^\star$  & 24& $ <3.6^\star$    \\  
 
\hline

\end{tabular}
\vskip 0.05in
$^\ast$From \citet{Neeleman17, Neeleman20}, but with the NUV SFR measured here, with updated HST-WFC3 calibrations.\\
$^\dagger$The field of the $z = 4.8166$ DLA towards J1054+1633 has not so far been searched for \cii\   emission.\\
$^\star$For FIR and NUV non-detections, we assume that the SFR$_{\rm FIR}$ and SFR$_{\rm NUV}$ values are equal to the $3\sigma$ upper limits when estimating the stacked SFR$_{\rm FIR}$, SFR$_{\rm NUV}$ and $\tau_{\rm depl}$ values. The possible error  from this assumption is a factor  $\lesssim 1.5$.
\end{table}

\begin{figure}
\centering
\includegraphics [scale=0.34] {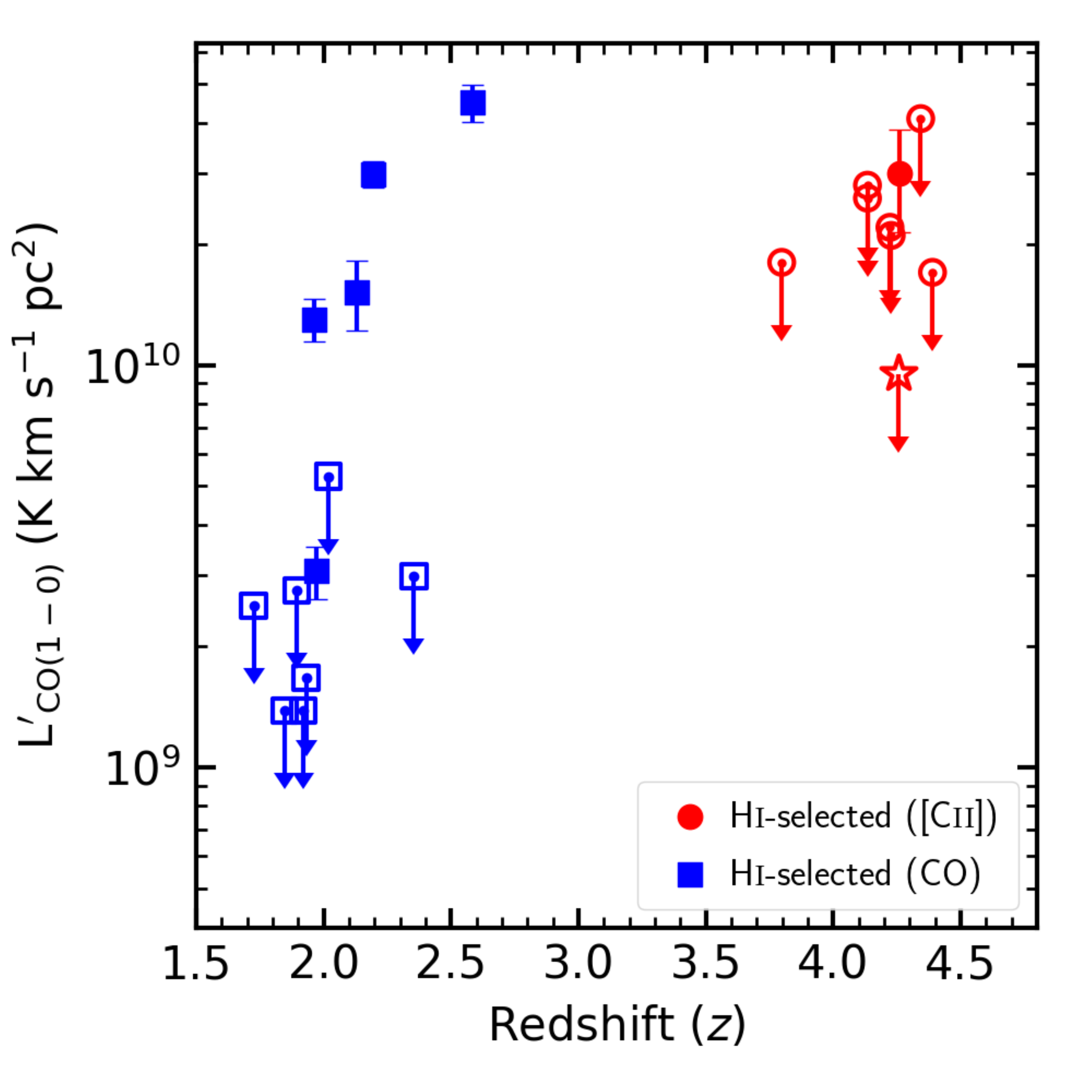}
\caption{The CO(1--0) line luminosity plotted against the DLA redshift, for the ALMA sample of \hi-selected galaxies at $z \approx 1.8-4.5$. The red circles represent the \hi-selected \cii-emitting galaxies  $z\approx 4$ (this work), and the blue squares represent the \hi-selected galaxies with CO searches at $z\approx 2$ \citep{Kanekar20}. Filled symbols indicate CO detections, while CO non-detections, and $3\sigma$ upper limits to the line luminosity, are indicated by open symbols and downward-pointing arrows. The red star represents the upper limit on the average CO(2--1) line luminosity of the six \hi-selected galaxies at $z \approx 4$. Five of the twelve \hi-selected galaxies at $z \approx 2$ are detected in CO emission, with four having high CO(1--0) line luminosities, while only one of the eight galaxies at $z \approx 4$ shows a CO detection.
\label{fig:redshift}}
\end{figure}

\section{Discussion}
\subsection{Molecular gas and dust in high-$z$ \hi-selected galaxies}
JVLA or ALMA searches for CO emission have now been carried out in more than 30 absorption-selected galaxies, at $z \approx 0.1 - 4.8$, with more than 15 detections of CO emission \citep[e.g.  ][]{Neeleman16,Neeleman18,Neeleman20,Moller18,Kanekar18,Kanekar20,Fynbo18,Klitsch19,Peroux19}. At $z < 1$ and $z \approx 4$, most of the \hi-selected galaxies had earlier been identified via their optical or \cii\ emission, while, at $z \approx 2$, the CO searches were themselves used to identify the \hi-selected galaxy. The CO searches have mostly targeted DLAs with high metallicities, typically in the top decile of the metallicity distribution at the DLA redshift. We emphasize that the evolution of DLA metallicity with redshift \citep{Rafelski12} implies that the targeted DLAs have very different metallicities at different redshifts, with a median metallicity [M/H]$_{\rm med} \approx -0.3$ at $z \lesssim 0.8$ and $z \approx 2$ \citep{Kanekar18,Kanekar20}, and of [M/H]$_{\rm med} \approx -1$ at $z \approx 4$ \citep[][this work]{Neeleman17, Neeleman19}.

\begin{figure}
\centering
\includegraphics [width = 0.32\textwidth] {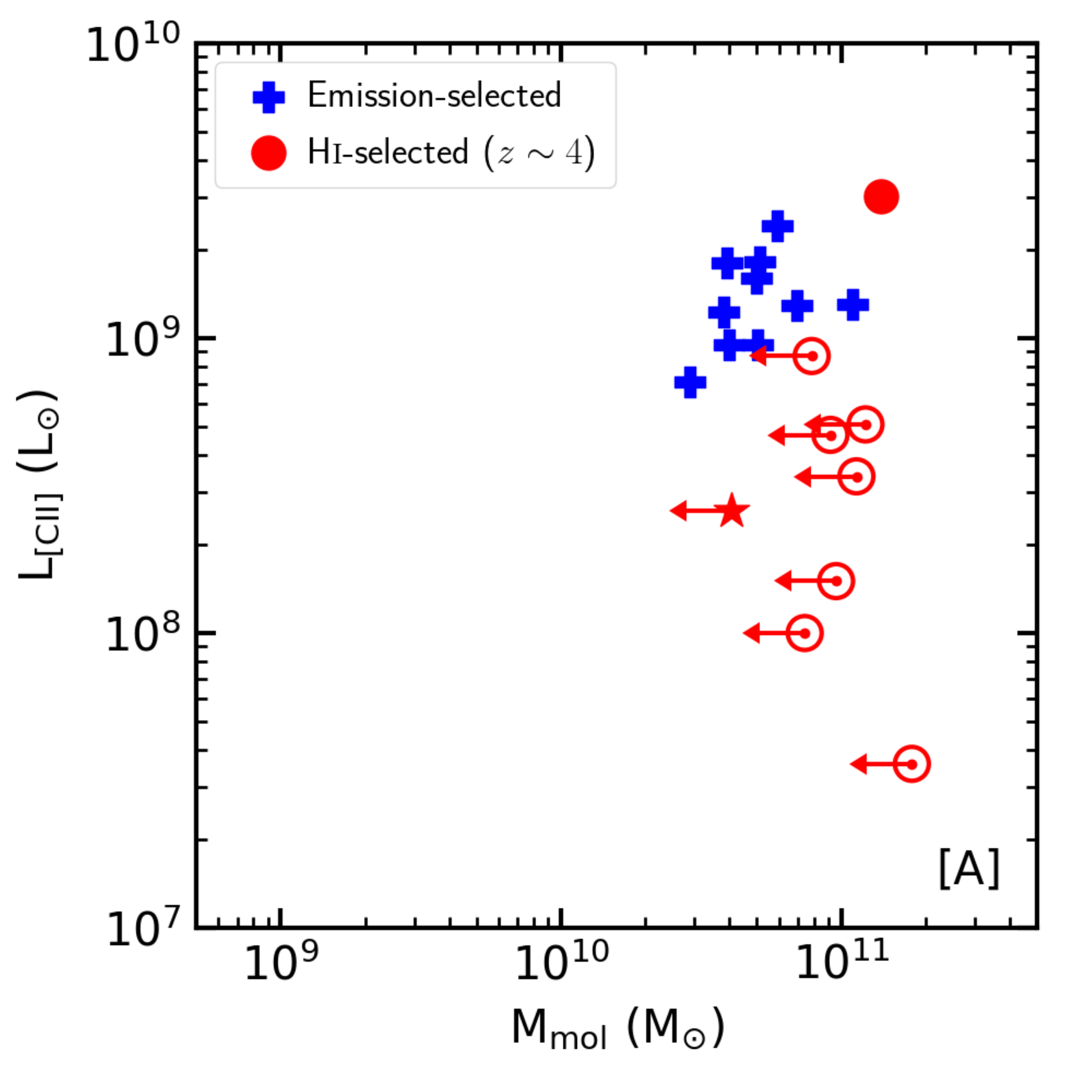}
\includegraphics [width = 0.32\textwidth] {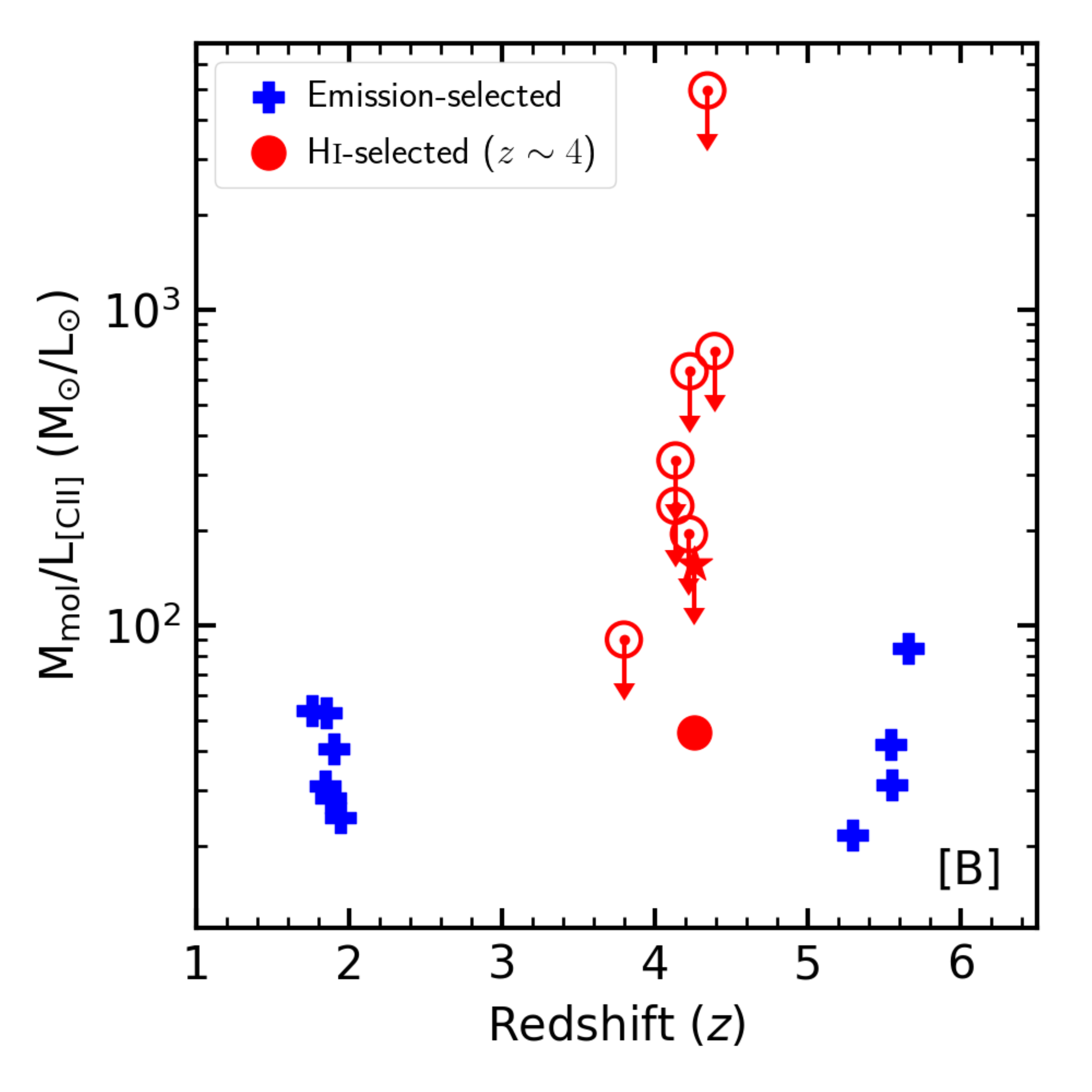}
\includegraphics [width = 0.32\textwidth] {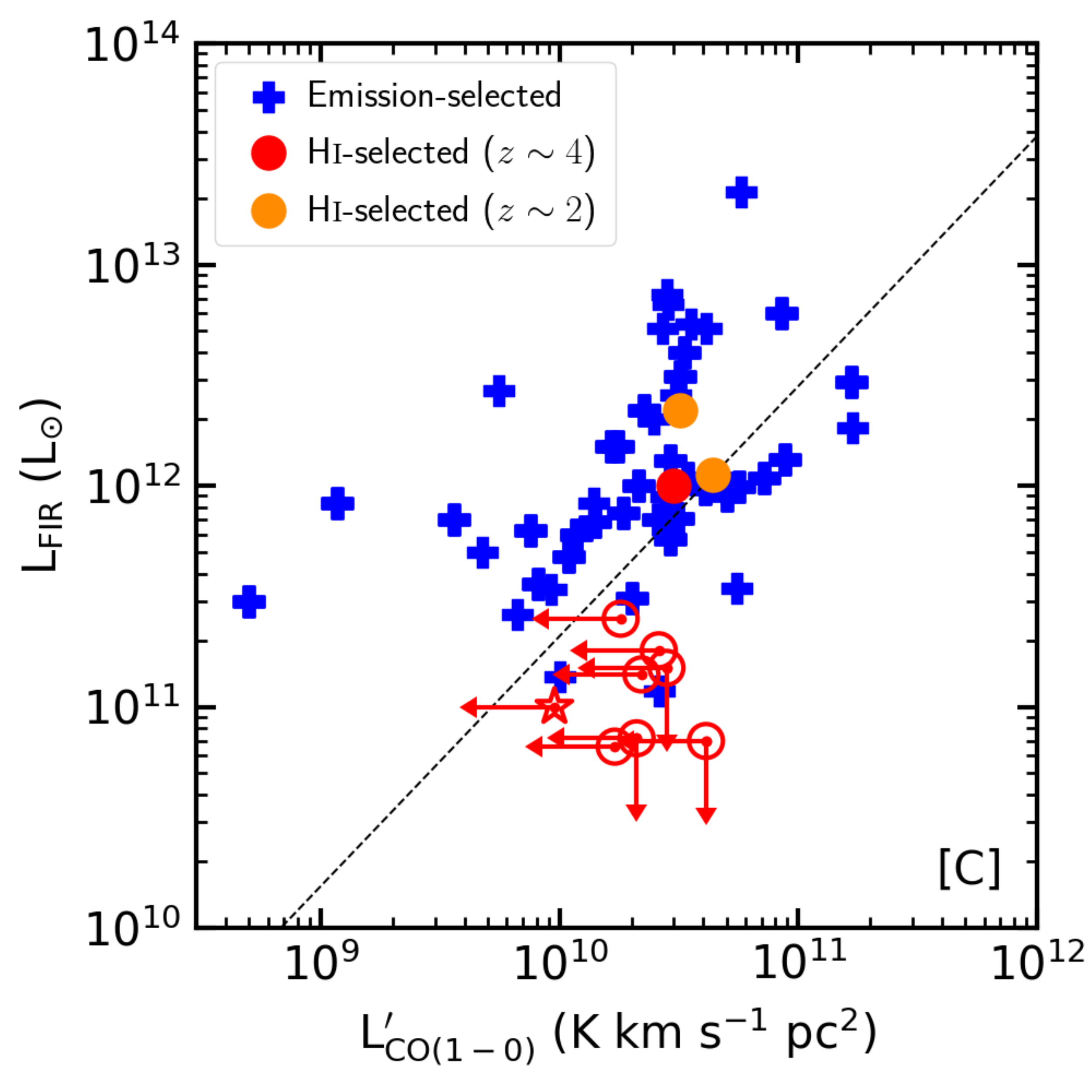}
\caption{[A]~Left panel: The \cii\ line luminosity, $\lcii$, plotted against the molecular gas mass, $\Mmol$, for the \hi-selected galaxies at $z \approx 4$ (filled and open red circles for CO detections and non-detections, respectively, with the ``CO stack'' shown as a red star), along with emission-selected star-forming galaxies at $z \approx 2$ and $z \approx 5$ \citep[blue pluses; ][]{Capak15, Zanella18}. The \hi-selected galaxies are seen to be consistent with the $\rm \lcii - \Mmol$ trend of emission-selected galaxies \citep{Zanella18}.
    [B]~Middle panel: The ratio $\Mmol/\lcii$ plotted against redshift, for the galaxies of the left panel. 
    [C]~Right panel: The FIR luminosity, $\rm L_{FIR}$, plotted against the CO(1--0) line luminosity, $\rm L'_{CO(1-0)}$ (both in logarithmic units), for the \hi-selected galaxies at $z \approx 4$ (red circles) and $z \approx 2$ (orange circles), and a comparison sample of high-$z$ emission-selected galaxies \citep[blue pluses; e.g. ][]{Carilli13,Zavadsky15, Magdis17, Pavesi19, Kaasinen19, Cassata20,Aravena20}. The dashed line indicates the fit to the  $\rm L_{FIR}-L'_{CO(1-0)}$ relation for the main-sequence galaxies of \citet{Genzel10}.
\label{fig:correlation_cii}}
\end{figure}

Fig.~\ref{fig:redshift} shows the CO(1$-$0) line luminosity $\rm L'_{CO(1-0)}$ plotted against absorber redshift for the \hi-selected galaxies at $z \approx 2$ and $z \approx 4$ \citep[][this work]{Kanekar20}. Four of the 12 high-metallicity absorption-selected galaxies at $z \approx 2$ have high CO(1$-$0) line luminosities, $\rm L'_{CO(1-0)} \gtrsim 10^{10}$~K~\kms~pc$^{2}$, while only one of the eight galaxies at $z \approx 4$ has a similarly high CO(1$-$0) line luminosity. CO emission surveys of deep fields have found evidence that the molecular gas mass density of the Universe increases from $z \approx 4$ to $z \approx 2$ \citep[e.g.][]{Riechers19,Decarli20}. Such surveys are sensitive to galaxies at the upper end of the metallicity distribution (as the higher $\aco$ of low-metallicity galaxies makes it  difficult to detect them in unbiased surveys), while our target DLA galaxies have been identified based on the high absorption metallicity, i.e. are likely to themselves have high metallicities. The higher CO line luminosity (and, presumably, molecular gas mass) in high-metallicity \hi-selected galaxies at lower redshifts is consistent with the above picture of an increasing molecular gas mass density with time.

Our $z \approx 4$ \hi-selected galaxies show strong \cii\ emission and, in most cases, detections of far-infrared (FIR) dust continuum emission. However, the JVLA observations yielded only upper limits on the CO line luminosity and hence, on the molecular gas mass. Stacking the CO(2--1) emission signals from six of the \hi-selected galaxies also did not yield a detection of CO emission, albeit with more stringent limits on the average CO line luminosity and the average molecular gas mass of the sample. The only $z \approx 4$ \hi-selected galaxy with a detection of CO emission thus remains DLA0817g1 at $z \approx 4.2603$, the Wolfe disk  \citep{Neeleman20}; this also has the highest \cii\ line luminosity and FIR luminosity of all the \hi-selected galaxies of our sample.

Fig.~\ref{fig:correlation_cii}[A] plots the \cii\ line luminosity of the \hi-selected galaxies against the molecular gas mass inferred from our JVLA studies; the eight \hi-selected galaxies are shown as red circles, while the average result from the CO(2--1) stacking is shown as a red star. For comparison, we have included high-redshift emission-selected galaxies \citep[main-sequence galaxies and Lyman-break galaxies, but excluding sub-mm galaxies; ][]{Capak15,Zanella18}; 
It is clear that DLA0817g1, the only galaxy of our sample with detections of both \cii\ and CO emission \citep{Neeleman20}, has a \cii\ line luminosity and a molecular gas mass consistent with those of the emission-selected galaxies. The seven  \hi-selected galaxies with upper limits on the molecular gas mass (as well as the upper limit obtained from the CO(2--1) stacking analysis) have systematically lower \cii\ line luminosities than DLA0817g1 and the emission-selected galaxies. The data on the \hi-selected galaxies are thus consistent with a correlation between the \cii\ line luminosity and the molecular gas mass \citep{Zanella18}. Fig.~\ref{fig:correlation_cii}[B] plots the ratio of the molecular gas mass to the \cii\ line luminosity against redshift for the same sample. The emission-selected galaxies at $z \approx 2$ and $z \approx 5$ have $\Mmol/\lcii$ values in the range  $10-100$, similar to the value in DLA0817g1. For the other \hi-selected galaxies, with CO non-detections, the upper limits on $\Mmol/\lcii$ are consistent with the above range.

Fig.~\ref{fig:correlation_cii}[C] plots the FIR luminosity $\rm L_{FIR}$ against the CO(1--0) line luminosity $\rm L'_{CO(1-0)}$ for the \hi-selected galaxies, again along with a comparison sample of emission-selected high-$z$ galaxies  \citep[including main-sequence galaxies, Lyman-break galaxies, and colour-selected galaxies, but excluding sub-mm galaxies; ][]{Carilli13,Zavadsky15, Magdis17, Pavesi19, Kaasinen19, Cassata20,Aravena20}. Earlier studies have found evidence of a correlation between $\rm L_{FIR}$ and $\rm L'_{CO(1-0)}$ in main-sequence galaxies \citep[e.g.][]{Daddi10,Genzel10}; the dashed line in the figure shows the fit of \citet{Genzel10} to the $\rm L_{FIR} - L'_{CO(1-0)}$ relation. Our single CO detection, and the upper limits on the CO line luminosity in the DLA galaxies are seen to be consistent with the correlation between $\rm L_{FIR}$ and $\rm L'_{CO(1-0)}$ obtained in high-$z$ main-sequence galaxies. 

It is clear from Fig.~\ref{fig:correlation_cii}[A] and [C] that the \hi-selected galaxies have typically lower \cii\ line and FIR continuum luminosities than those of the emission-selected samples. Our detection of CO emission in DLA0817g1 and the upper limits on the CO(1--0) line luminosity in the other \hi-selected galaxies, with implied modest molecular gas masses, are consistent with the trends from the more luminous and massive emission-selected population. Our CO results are thus consistent with the null hypothesis that the \hi-selected galaxies are normal main-sequence galaxies, selected by their absorption properties.

\subsection{Stellar properties of high-$z$ \hi-selected galaxies}

\begin{figure}
\centering
\includegraphics[scale=0.34]{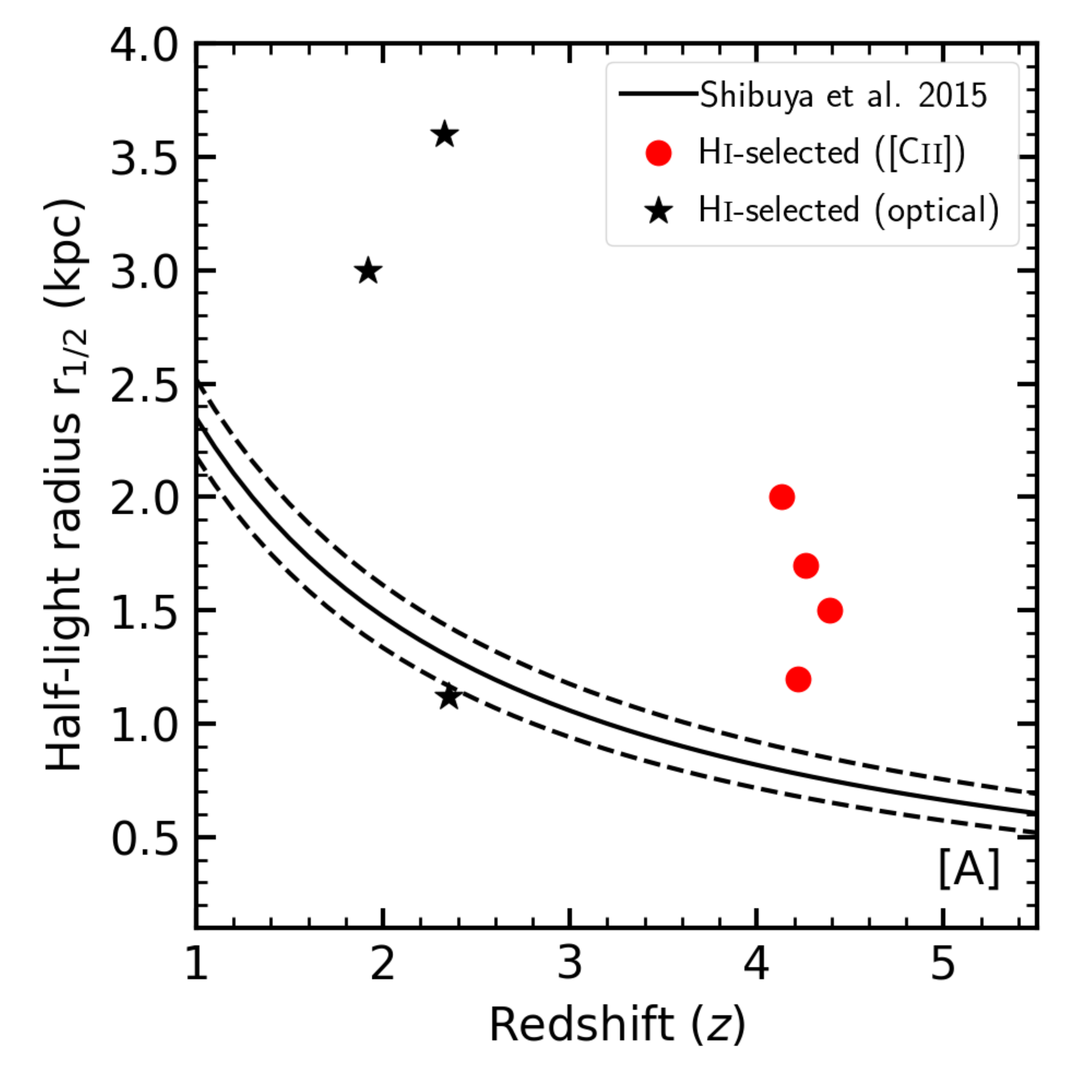}
\includegraphics[scale=0.34]{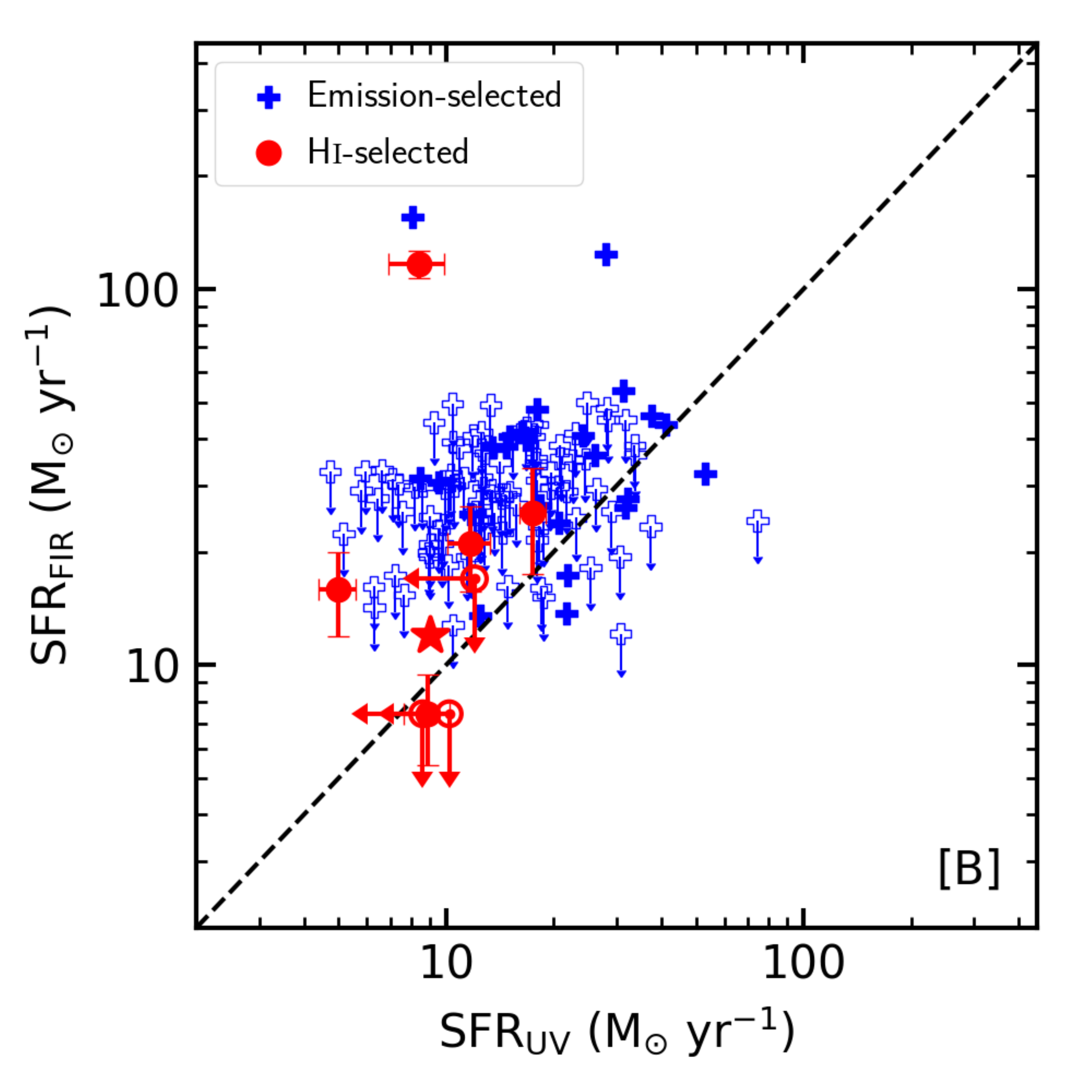}
\caption{[A]~Left panel: The half-light radius of the NUV or H$\alpha$ emission (kpc) plotted as a function of galaxy redshift; the \hi-selected galaxies of this paper are shown as filled red circles, while \hi-selected galaxies at $z \approx 2$, identified via optical or near-IR spectroscopy,  are shown as black stars \citep{Moller02, Krogager13, Bouche13}. The solid black curve shows the redshift dependence of the half-light radius for emission-selected galaxies with $L_{\rm UV}$ = 0.1--1 $L^{\ast}_{z=3}$ \citep{Shibuya15}, with the dashed curves showing the $\pm 1\sigma$ spread around the relation. Most of the \hi-selected galaxies are seen to have NUV sizes larger than those expected from the relation of \citet{Shibuya15}. [B]~Right panel: The total SFR (SFR$_{\rm FIR}$) plotted against the dust-unobscured SFR (SFR$_{\rm UV}$) for the \hi-selected galaxies \citep[red circles;][this work]{Neeleman17, Neeleman19} and emission-selected galaxies from the ALPINE \cii\ survey at $z \approx 4.4-5.8$ \citep[blue plus signs;][]{Bethermin20, Faisst20}. Filled symbols represent SFRs inferred from detections of both $160\mu$m continuum and NUV/FUV emission, while open symbols indicate $3\sigma$ upper limits on one or both SFRs. The dashed line indicates the case for zero dust obscuration, i.e. SFR$_{\rm UV} = $~SFR$_{\rm FIR}$.
\label{fig:NUV}}
\end{figure}

Over the last 25 years, there have been a number of HST and ground-based searches for the stellar counterparts of high-$z$ \hi-selected galaxies via their rest-frame UV or optical continuum, or their H$\alpha$ or Lyman-$\alpha$ emission \citep[e.g.][]{Warren01,Moller02,Moller04,Peroux11,Peroux12,Fynbo13,Fumagalli15,Hartoog15,Wang15,Krogager13,Krogager16,Mackenzie19,Ranjan20,Rhodin20}. Only around 20 galaxy counterparts have been identified via such searches, all at $z \lesssim 2.5$, with $\approx 10$ having detections of the stellar continuum \citep[e.g.][]{Krogager17,Rhodin20}. Our HST-WFC3 imaging has pushed such studies to the highest redshifts till date, yielding the first detections of the stellar emission of \hi-selected galaxies at $z \gtrsim 4$.

The half-light radii of the NUV emission of the 5 \hi-selected galaxies with NUV detections lie in the range $1.2-3.1$~kpc (see Table~\ref{tab:HST}), with a median half-light radius of 1.7~kpc. Interestingly, the median NUV half-light radius is  higher than that of the ALPINE main-sequence galaxies at $z \approx 4.4-5.8$ \citep[$r_e \approx 0.96$~kpc; ][]{Fujimoto20}. Fig.~\ref{fig:NUV}[A] plots the half-light radius of the \hi-selected galaxies at $z \approx 4$ (red circles\footnote{We exclude DLA1201g1, where the large measured NUV half-light radius includes contributions from both members of the merging system \citep{Prochaska19}.}) and $z \approx 2$ (black stars) \citep{Moller02, Krogager13, Bouche13} against galaxy redshift, along with the NUV size-redshift relation of \citet{Shibuya15} for emission-selected galaxies (black curve). We find that all the \cii-emitting \hi-selected galaxies at $z \approx 4$ and two of the three systems at $z \approx 2$ have half-light radii larger than expected from the size-redshift relation of emission-selected galaxies. 
This might suggest that the \hi-selection is identifying galaxies that are more spatially extended than emission-selected galaxies at the same redshift. 

Further, the spatial extents of the \cii\ emission and the NUV emission in Fig.~\ref{fig:HST} are very similar; this is especially clear for DLA1201g1 and DLA0817g1, which have high-resolution \cii\ images. The situation in the \hi-selected galaxies thus appears different from that in emission-selected galaxies at similar redshifts: \citet{Fujimoto20} find that the spatial extent of the \cii\ emission is $2-3$ times larger than that of the UV emission in the ALPINE sample of main-sequence galaxies at $z \approx 4.4-5.8$. We also find no evidence for spatial offsets between the NUV emission and the \cii\ emission, as has been earlier seen in emission-selected samples at $z \gtrsim 4$ \citep[e.g.][]{Maiolino15,Carniani18,Ginolfi20}.

In Fig.~\ref{fig:NUV}[B], we plot the total SFR (SFR$_{\rm FIR}$) against the dust-unobscured SFR (SFR$_{\rm UV}$) for the \hi-selected galaxies at $z \approx 4$ \citep[red circles][this work]{Neeleman17, Neeleman19}. For comparison, we also plot the same quantities for the main-sequence galaxies at $z = 4.4-5.8$ from the ALPINE \cii\ survey \citep[blue pluses; ][]{Bethermin20, Faisst20}. We estimated SFR$_{\rm UV}$ for the ALPINE galaxies from the FUV magnitudes listed in the ALPINE catalogue \citep{Bethermin20}, using the relation of \citet{Kennicutt12} and a Chabrier initial mass function to convert from FUV luminosity to SFR. The SFR$_{\rm FIR}$ for the ALPINE galaxies was determined from their 155$\mu$m luminosity, following the procedure of \citet{Neeleman19}. The SFRs of the \cii-emitting, \hi-selected galaxies appear consistent with those of the main-sequence ALPINE galaxies. We note that the NUV SFRs of the four new HST-WFC3 detections in the \hi-selected galaxies are consistent with, although typically slightly lower than, the total SFRs inferred from the ALMA $160\mu$m continua. This is very unlike the situation in the Wolfe disk (DLA0817g1), where the FIR SFR is nearly fifteen times larger than the NUV SFR \citep[][this work]{Neeleman20}. Further, all \hi-selected galaxies that are not detected in their NUV stellar emission are also not detected in the 160$\mu$m continuum, indicating that they have low total SFRs. This indicates that most of the \cii-emitting \hi-selected galaxies at $z \approx 4$ do not contain large amounts of  dust, which would obscure the NUV emission and result in an NUV SFR that is significantly lower than the total SFR.

We also estimated the SFRs of the \hi-selected galaxies using the \ci-SFR relation obtained by \citet{Schaerer20} for the ALPINE sample. This yields SFR$_{\scriptsize \textrm{[C} \textsc{ii} \textrm{]}}$ values in the range $3-282\ \Msun$~yr$^{-1}$ (see Table \ref{tab:derived}). For most of our objects, the SFR estimates from the ALPINE \ci-SFR relation are higher than the total SFRs inferred from the FIR luminosity, in some cases by a factor of $\approx 3$.

Finally, the \cii ~and FIR $160\mu$m contours in Figs.~\ref{fig:HST} and \ref{fig:HSTdust} for DLA1054g1 fall between two galaxies in the HST-WFC3 image. We have identified the fainter NUV-emitting galaxy as the counterpart as its NUV SFR is more consistent with the FIR SFR. If the associated galaxy is the brighter galaxy, then its NUV SFR would be $\approx  54\ \Msun$~yr$^{-1}$, significantly higher than that derived from the ALMA dust $160\mu$m continuum. Either way, the Wolfe disk remains an outlier as an \hi-selected, dusty high-redshift disk galaxy.

\subsection{Molecular gas depletion times of \hi-selected galaxies}

The molecular gas depletion timescale $\tau_{\rm depl} \equiv \Mmol/{\rm SFR}$ quantifies the time in which the molecular gas of a galaxy would be entirely converted to stars, if the galaxy continued to form stars at its present SFR, without further formation of molecular gas. \citet{Tacconi20} find that $\tau_{\rm depl}$ depends primarily on redshift and the distance of the galaxy from the main-sequence relation at its redshift \citep[see also][]{Genzel15,Tacconi18}: for example, at a given redshift, starburst galaxies have systematically lower values of $\tau_{\rm depl}$ than main-sequence galaxies. Estimates of $\tau_{\rm depl}$ in galaxies can thus be used to infer whether the galaxies are likely to lie on the main sequence (or above or below it). The last column of Table~\ref{tab:derived} lists the molecular gas depletion times (or limits to $\tau_{\rm depl}$) of the \hi-selected galaxies, using the SFR estimates from the FIR luminosity \citep{Neeleman17,Neeleman19}. It can be seen that $\tau_{\rm depl} \approx 1.2$~Gyr for DLA0817g1, $\tau_{\rm depl} < 3.3$~Gyr for DLA1201g1, and $\tau_{\rm depl} < 3.6$~Gyr for the stacked CO(2--1) emission. The single measurement of, and the upper limits on, $\tau_{\rm depl}$ in \hi-selected galaxies at $z \approx 4$ are in good agreement with the range of $\tau_{\rm depl}$ values obtained by \citet[][see their Fig.~3]{Tacconi20} for main-sequence galaxies at $z \approx 4$. \citet{Kanekar18} found that \hi-selected galaxies at intermediate redshifts, $z \approx 0.7$, have far larger values of $\tau_{\rm depl}$, $\approx 10$~Gyr, than main-sequence galaxies. Unlike their counterparts at $z \approx 0.7$, the \hi-selected galaxies at $z \approx 4$ show no evidence for large molecular gas reservoirs and low SFRs; their properties are consistent with those of main-sequence galaxies at similar redshifts.

\section{Summary}
We have used the JVLA to search for redshifted CO(1--0) or CO(2--1) emission from seven \cii-emitting galaxies associated with high-metallicity ([M/H]~$\geq -1.3$) DLAs at $z \approx 4$, and from the field of a low-metallicity ([M/H]~$=-2.47$) DLA at $z \approx 4.8$. Our non-detections of CO emission yield upper limits of $\rm \Mmol < (7.4 - 17.9) \times 10^{10} \times (\aco/4.36)  \; \Msun$ on the molecular gas mass of the DLA galaxies, assuming sub-thermal excitation of the J $=2$ level. By stacking the CO(2--1) emission signals of six of the \cii\ emitters in the image plane, we obtain an upper limit of $\rm \Mmol < 4.1 \times 10^{10} \times (\aco/4.36)  \; \Msun$ on the average molecular gas content of these galaxies. 

In addition, we have used WFC3 onboard HST to image the above \hi-selected galaxies at $z \approx 4$ in the rest-frame NUV stellar continuum. We obtain four new detections of NUV emission, resulting in NUV SFRs of $\approx (5.0 - 17.5) \, \Msun$~yr$^{-1}$, and three upper limits, yielding SFR$_{\rm NUV} < (8.6-12.0) \, \Msun$~yr$^{-1}$. We also re-analysed the HST-WFC3 data on the $z \approx 4.2603$ \cii-emitting galaxy towards J0817+1351 \citep{Neeleman20} using new HST-WFC3 calibrations at a smaller plate-scale, and obtain a lower NUV SFR, by a factor of $\approx 2$. For the four HST-WFC3 detections, the NUV SFRs are in reasonable agreement with the SFRs estimated from the FIR $160\mu$m continuum emission, suggesting that most of the $z \approx 4$ \hi-selected galaxies do not suffer from significant dust obscuration. For the five NUV detections, the half-light radius of the NUV emission is $\approx 1.2-3.1$~kpc, with a median half-light radius of $1.7$~kpc, larger than that of the ALPINE main-sequence galaxies at $z \approx 4.4-5.8$, and also above the size-redshift relation of emission-selected galaxies. The spatial extents of the NUV and \cii\ emission are similar for the \hi-selected galaxies; this too is different from the ALPINE main-sequence galaxies, for which the \cii\ emission is spatially more extended, by a factor of $\approx 2-3$, than the NUV emission.

We find that the \cii\ line luminosity, FIR luminosity and molecular gas mass of \hi-selected galaxies at $z \approx 4$ are consistent with the correlations between \cii\ line luminosity and molecular gas mass, and between FIR luminosity and CO(1--0) line luminosity, that have earlier been identified in emission-selected galaxies at high redshifts. Our upper limits on the molecular gas mass allow us to measure, or place limits on, the molecular gas depletion time, $\tau_{\rm depl}$ of the \hi-selected galaxies. We obtain $\tau_{\rm depl} = 1.2$~Gyr for DLA0817g1 \citep{Neeleman20}, $\tau_{\rm depl} < 3.3$~Gyr for DLA1201g1, and $\tau_{\rm depl} < 3.6$~Gyr from the stacked CO(2--1) emission for the other \hi-selected galaxies, all consistent with the range of $\tau_{\rm depl}$ values in main-sequence galaxies at $z \approx 4$. We thus find no evidence for large molecular gas reservoirs in the \hi-selected galaxies at $z \approx 4$ such as those seen in absorption-selected galaxies at intermediate redshifts, $z \approx 0.7$. 

This is the first study of the molecular, \cii, NUV, and dust characteristics of \hi-selected galaxies at any redshift. We find that the \hi-selected galaxies are consistent with most of the trends seen in high-redshift emission-selected samples, extending these trends to lower-luminosity galaxies. However, the NUV emission in \hi-selected galaxies appears to be more spatially-extended than that in emission-selected galaxies at similar redshifts. Also, their \cii\ and NUV emission have similar spatial extents, again unlike the situation in emission-selected galaxies. Further such studies of \hi-selected galaxies are critical to elucidate the similarities and differences between them and emission-selected galaxies, and obtain a more complete picture of high-redshift galaxy formation and evolution.

\acknowledgements
The National Radio Astronomy Observatory is a facility of the National Science Foundation operated under cooperative agreement by Associated Universities, Inc. Support for Program numbers 15410 and 15882 were provided by NASA through grants from the Space Telescope Science Institute, which is operated by the Association of Universities for Research in Astronomy, Incorporated, under NASA contract NAS5-26555.  BK and NK  acknowledge the Department of Atomic Energy for funding support, under project 12-R\&D-TFR-5.02-0700. NK also acknowledges support from the Department of Science and Technology via a Swarnajayanti Fellowship (DST/SJF/PSA-01/2012-13). M.N. acknowledges support from ERC Advanced grant 740246 (Cosmic\texttt{\char`_}Gas).

\bibliographystyle{aasjournal}

\end{document}